\documentclass[a4paper,fleqn]{cas-sc}

\usepackage[round, sort, numbers]{natbib}

\usepackage{float}
\usepackage{caption}

\def\tsc#1{\csdef{#1}{\textsc{\lowercase{#1}}\xspace}}
\tsc{WGM}
\tsc{QE}
\tsc{EP}
\tsc{PMS}
\tsc{BEC}
\tsc{DE}

\begin{document}
\let\WriteBookmarks\relax
\def\floatpagepagefraction{1}
\def\textpagefraction{.001}

\shorttitle{Survey of Machine Learning Based Intrusion Detection Methods for Internet of Medical Things}

\shortauthors{A. Si-Ahmed et~al.}

\title [mode = title]{Survey of Machine Learning Based Intrusion Detection Methods for Internet of Medical Things}                      


%

\author[1,4]{Ayoub Si-Ahmed}[orcid=0000-0002-2296-2783]
\ead{si_ahmed.ayoub@etu.univ-blida.dz, ayoub.siahmed@proxylan.dz, ayoub.siahmed.94@gmail.com}
\author[2]{Mohammed Ali Al-Garadi}[orcid=0000-0001-5204-2980]
\ead{m.a.al-garadi@emory.edu}
\author[3]{Narhimene Boustia}[orcid=0000-0002-4258-4468]
\ead{nboustia@gmail.com, nboustia@univ-blida.dz }

\address[1]{Blida 1 University, LRDSI Laboratory, Blida, B.P 270, Algeria}
\address[2]{Emory University, School of Medicine, Department of Biomedical Informatics, Atlanta, GA 30322, USA}
\address[3]{Blida 1 University, SIIR/LRDSI (Blida1) \& RCR/RIIMA (USTHB) Laboratory, Blida, B.P 270, Algeria}
\address[4]{PROXYLAN SPA/Subsdiary of CERIST, Algeria, 16028, Algeria}




     
\begin{abstract}
The Internet of Medical Things (IoMT) has revolutionized the healthcare industry by enabling physiological data collection using sensors, which are transmitted to remote servers for continuous analysis by physicians and healthcare professionals. This technology offers numerous benefits, including early disease detection and automatic medication for patients with chronic illnesses.
However, IoMT technology also presents significant security risks, such as violating patient privacy or exposing sensitive data to interception attacks due to wireless communication, which could be fatal for the patient. 
Additionally, traditional security measures, such as cryptography, are challenging to implement in medical equipment due to the heterogeneous communication and their limited computation, storage, and energy capacity. These protection methods are also ineffective against new and zero-day attacks.
It is essential to adopt robust security measures to ensure data integrity, confidentiality, and availability during data collection, transmission, storage, and processing. In this context, using Intrusion Detection Systems (IDS) based on Machine Learning (ML) can bring a complementary security solution adapted to the unique characteristics of IoMT systems. 
Therefore, this paper investigates how IDS based on ML can address security and privacy issues in IoMT systems. First, the generic three-layer architecture of IoMT is provided, and the security requirements of IoMT systems are outlined. Then, the various threats that can affect IoMT security are identified, and the advantages, disadvantages, methods, and datasets used in each solution based on ML at the three layers that make up IoMT are presented. Finally, the paper discusses the challenges and limitations of applying IDS based on ML at each layer of IoMT, which can serve as a future research direction.
\end{abstract}



\maketitle

\section{Introduction}
The Internet of Things (IoT) represents the fourth industrial revolution, in which devices and systems are connected to enable communication and data exchange. The first three revolutions focused on agriculture, industry, and information technology. Kevin Ashton, a British technology pioneer, coined the term IoT in 1999 \cite{suresh2014state} and defined it as a network of physical objects embedded with electronics, software, sensors, and network connectivity, allowing these objects to collect and exchange data, often using the internet \cite{wikipedia_2022}. The use of sensors enables machine-to-machine communication, which allows for the exchange of data and information without the need for human intervention. This technology can potentially revolutionize various sectors, including healthcare, transportation, and manufacturing.

IoT is experiencing considerable development and is estimated to reach more than 24.1 billion devices worldwide in 2030, representing about four devices per person  \cite{tollefson_tollefson}, which can be explained by the many possible applications that offer IoT. Among these applications is the Internet of Medical Things (IoMT). The IoMT uses sensors that are either wearable or implemented in the human body to collect health data. Then these data are sent to a remote server to be analyzed using Artificial Intelligence (AI) assisted by the physicians.

Healthcare has experienced many evolutions in digitalizing health data, as explained in \cite{SiemensHealthcare2019} and summarized in this paragraph. The evolution of healthcare began in 1990 with the advent of what is now known as healthcare 1.0. This initial phase involved digitizing medical records, where doctors started entering patient notes into computers. These digital records were then managed and stored by specialized systems such as the Picture Archiving and Communication System and Radiology Information System.
The introduction of Healthcare 2.0 saw hospitals adopting systems that integrate and manage the digital data stored on doctors' laptops. Then there was healthcare 3.0, which consists of compiling, and grouping all patient data into an Electronic Health Record (EHR), providing individuals with complete access to their health data and history. Healthcare 4.0 is currently underway, enabled by artificial intelligence, data provided by doctors, imaging centers, equipment, and sensors implemented or worn by patients. This evolution allows doctors and medical staff to make more accurate diagnoses and better treatment decisions and allows hospital managers to control costs.

The IoMT is a rapidly growing field leveraged in various health-related applications. One of its key advantages is its ability to provide real-time and continuous patient health monitoring. This technology has proven particularly effective in managing chronic conditions, such as diabetes, where insulin pumps can automatically inject insulin to regulate glucose levels. Similarly, pacemakers, which send electric shocks to the heart when an abnormality is detected, have been used to manage heart disease. Deep Brain Implants (DBI), another medical device enabled by the IoMT, provide electrical stimulation to the brain to treat neurological disorders like Parkinson's. Beyond treating specific conditions, the IoMT has been used for fall detection in older adults and performance measurement for athletes. In addition, it could improve access to healthcare in rural areas that lack medical infrastructure.

Moreover, the IoMT produces detailed and accurate medical data, increasing treatment efficiency, reducing medical errors, and identifying diseases in their early stages. This can lead to quicker treatment and improved patient outcomes. By transforming healthcare from curative to preventive, the quality of care for patients is significantly improved, and stakeholders like insurance companies and pharmacies benefit as well.
Another advantage of IoMT is the remote access to medical data by nurses and patients' families, eliminating the need for regular hospital visits, reducing costs, saving hospital resources, and containing the spread of COVID-19. Additionally, patients can receive care in the comfort of their own home. Overall, the IoMT is a remarkable technological advancement that has the potential to significantly improve healthcare services, benefiting both patients and healthcare providers.

While the IoMT has the potential to revolutionize healthcare through real-time patient monitoring, its adoption faces significant barriers due to security and privacy risks. The use of wireless communication to transmit data from sensors to remote servers creates security vulnerabilities that compromise the data’s confidentiality, integrity, and availability, with potentially fatal consequences for patients if incorrect treatments are administered. Several studies  \cite{rathore2019novel, rathore2017dlrt, kintzlinger2020cardiwall} have demonstrated that even critical medical devices like Implantable Cardioverter-Defibrillators (ICD), DBIs, and insulin pumps can be hacked. Furthermore, because IoMT data is highly sensitive and personal, unauthorized access or disclosure could result in severe patient privacy violations. Such incidents have occurred in recent years, for example, the Singapore health service data breach that affected 1.5 million patients \cite{tollefson_tollefson} and the ransomware attack on a French hospital’s system \cite{tollefson_tollefson}. Given these risks, it is essential to prioritize the security and privacy of IoMT data in order to ensure that patient safety is not compromised.

Securing IoMT presents unique challenges that traditional security methods cannot efficiently address. One such challenge is the low computational and storage capacity of IoMT sensors, which makes it difficult to apply resource-intensive security measures. In addition, designing a secure system for IoMT must consider caregivers' need for quick access during medical emergencies, as prompt response times can mean the difference between life and death. Moreover, the potential for third-party devices to control and upgrade IoMT equipment increases the risk of security breaches, necessitating the implementation of lightweight security solutions. However, since the attack vectors against IoMT systems constantly evolve, these solutions must be regularly revised and updated to ensure their efficacy against zero-day and new attacks. 

The following advantages motivated our choice to investigate the use of ML for IoMT intrusion detection: 

\begin{enumerate}[I]
\itemsep=0pt
\item Machine Learning can give intelligence to the IoMT system and be more suitable to their security requirements. Therefore, these methods can be more effective in emergencies than traditional access control methods.
\item IoMT and IoT systems generate a large amount of data that can be considered as big data due to their velocity, variety, and volume. These data are valuable sources for security because they can be used to learn normal behavior and detect abnormal behavior at an early stage, limiting the attack's damage.
\item Deep Learning (DL) algorithms can extract the relevant attributes to perform the classification automatically, which eliminates the necessary extraction process required in traditional ML methods and therefore offer an end-to-end security model \cite{mclaughlin2017deep}.
\item ML methods can detect zero-day attacks and new vulnerabilities, which threat signature-based methods cannot.
\end{enumerate}  

In light of these different advantages that can offer the use of ML in a security application, this study investigates its application for the IoMT by answering the following questions:

\begin{enumerate}[A]
\itemsep=0pt
    \item What is the generic architecture of an IoMT system, its security requirements, and the security threats that can affect it?
    \item What are the different ML-based solutions proposed at the different layers of the IoMT?
    \item What are the advantages and disadvantages of these different solutions? 
    \item What datasets are used to train and test the ML model at the different layers of IoMT?
\end{enumerate}

The paper is divided into several sections to address the topic of IoMT security comprehensively. In Section \ref{ARCHITECTURE OF IOMT}, the different layers of IoMT architecture are described. Section \ref{SECURITY REQUIREMENTS OF IOMT} discusses the various requirements that are necessary to ensure IoMT security. Section \ref{IOMT SECURITY THREAT} presents diverse threats that can undermine IoMT security, detailing the possible attacks that can affect each layer of the system, the types of attacks, attack environments, and the adversary’s motivation to conduct an attack. The paper also highlights the significance of IDS in Section \ref{INTRUSION DETECTION SYSTEM}, while Section \ref{MACHINE LEARNING OVERVIEW} presents an overview of the ML technique. Section \ref{REVIEW OF MACHINE LEARNING APPLICATIONS FOR IOMT SECURITY} discusses the state-of-the-art approaches proposed in different layers of the system, highlighting the benefits, drawbacks, and datasets used in each solution. Section \ref{CHALLENGES AND LIMITATIONS} sheds light on the limitations and challenges in using IDS based on ML for each layer of the IoMT system. Finally, the paper concludes in Section \ref{CONCLUSION}.

\begin{figure*}[!ht]
\centering
\includegraphics[width=\textwidth]{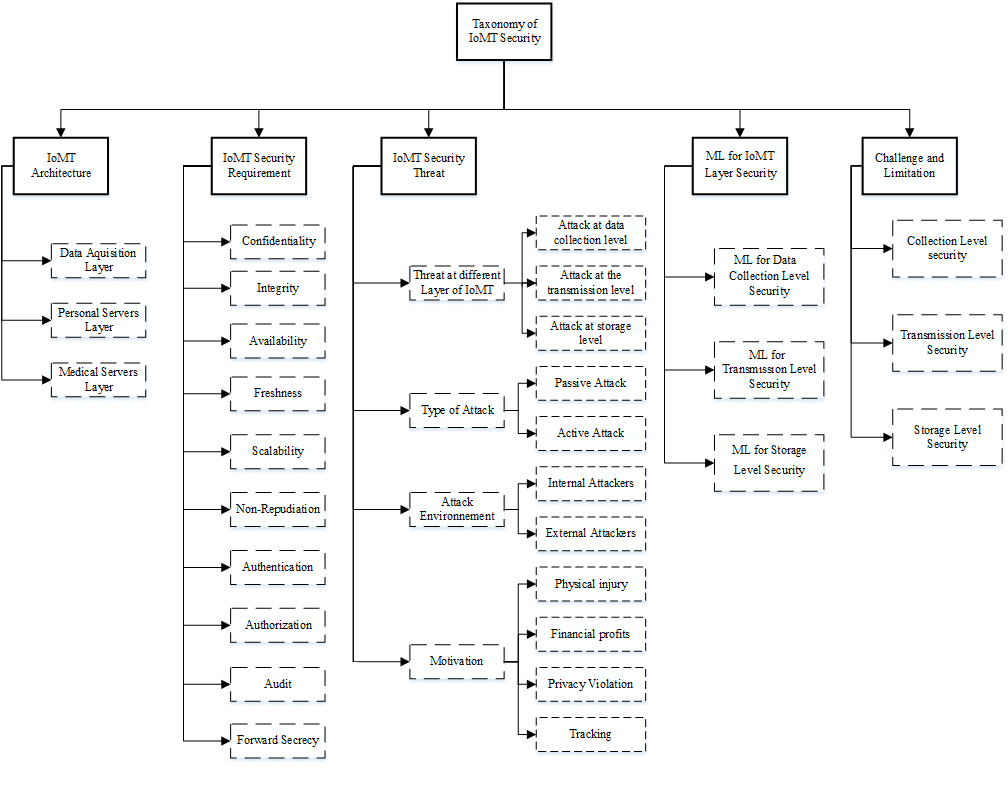}
\caption{Thematic Taxonomy of ML for IoMT Security}
\label{FIG:1}
\end{figure*}

Figure \ref{FIG:1} shows the various components explored in this review paper. 

\begin{table}[cols=2,pos=H]
\caption{List of abbreviations}\label{tbl6}
\begin{tabular*}{\tblwidth}{m{3cm} m{10cm} }
\toprule
IoMT & Internet of Medical Things\\
ML & Machine Learning\\
IDS & Intrusion Detection System\\
AI & Artificial Intelligence \\
EHR & Electronic Health Record\\
ICD & Implantable Cardioverter-Defibrillator\\
DL & Deep Learning\\
WBAN & Wireless Body Area Networks\\
DBI & Deep Brain Implants\\
EMR & Electronic Medical Record\\
DoS & Denial of Service\\
DDoS & Distributed Denial of Service\\
IMD & Implementable Medical Device\\
TPR & True Positive Rate\\
FPR & False Positive Rate\\
AUC & Area Under the Curve\\
MLP & Multi-Layer Perceptron\\
LPU & Local Processing Unit\\
MM & Markov Model\\
NFC & Near Field Communication \\
BLE & Bluetooth Low Energy \\
GSM & Global System for Mobile Communications \\
EEG & Electroencephalogram \\
PG & Pulse Generator \\
RMSE & Root Mean Squere Error\\
IHSS & Intelligent Healthcare Security System\\
MITM & Man-In-The-Middle\\
OS-ELM & Online Sequential Extreme Learning Machine\\
EOS-ELM & Ensemble of Online Sequential Extreme Learning Machine \\
ELM & Extreme Learning Machine\\
SVM & Support Vector Machine\\
DT & Decision Tree\\
RF & Random Forest\\
NB & Naive Bayes\\
ANN & Artificial Neural Network\\
KNN & k-Nearest Neighbor\\
SOM & Self-Organizing Map\\
CNN & Convolutional Neural Network\\
IoT & Internet of Things\\
ECG & Electrocardiogram\\
LSTM & Long Short-Term Memory\\
MLP & Multi-Layer Perceptron\\
IPG & Implementable Pulse Generation\\
FAR & False Alarm Rate \\
RTV & Rest Tremor Velocity\\
PMD & Personal Medical Device\\
DNN & Deep Neural Network\\
PCA & Principal Component Analysis\\
GWO & Grey Wolf Optimization\\
SDN & Software-Defined Networking\\
LR & Logistic Regression\\
FCM & Fuzzy C-Means Clustering\\
HTTPS & HyperText Transfer Protocol Secure\\
EMA & Exponential Moving Average\\

\bottomrule
\multicolumn{1}{r}{\footnotesize\itshape Continue on the next page}
\end{tabular*}
\end{table}

\begin{table}[cols=2,pos=H]
\caption{List of abbreviations}\label{tbl6_1}
\begin{tabular*}{\tblwidth}{m{3cm} m{10cm} }
\toprule
FPGA & Field Programmable Gate Arrays\\
Wi-Fi & Wireless Fidelity\\
CPU & Central Processing Unit\\
HFL & Hierarchical Federated Learning\\
RNN & Recurrent Neural Network\\
GRU & Gated Recurrent Unit\\
XAI & Explainable AI\\
\bottomrule
\end{tabular*}
\end{table}

\section{RELATED WORK}
Many review articles address different aspects of security in an IoMT system and include ML to ensure security in these systems. To the best of our knowledge, most of them do not focus on using ML as a method to ensure security in IoMT, and they just mentioned it briefly. Table \ref{tbl1} presents the different review papers that have discussed security in the IoMT, mentioning our main contribution compared with them. 

\emph{H.Rathore, et al.} explored issues, security risks relating to the privacy and safety of medical equipment, and solutions in \cite{rathore2017review}, including anomaly detection based on ML.

\emph{M.Hussain, et al.} in \cite{hussain2019authentication} reviewed different authentication schemes and classified them according to their type. They give advantages, disadvantages, and the ability to resist different attacks. They also categorized authentication mechanisms based on advanced methodologies, such as game theory and ML.

The work of \emph{M.Wazid, et al.} in \cite{wazid2019iomt} covered a variety of malware attacks against IoMT systems, targeting security criteria, namely confidentiality, integrity, authenticity, and availability of data. The current security approach strategy has generally emphasized key management and intrusion detection using different methods such as ML, authentication, and access control.

In the work conducted by \emph{A.I.Newaz, et al.} in \cite{newaz2021survey}, they discussed security and privacy in healthcare by presenting a detailed survey of possible attacks, their impact, and solutions proposed in the literature, including solutions based on ML along with their limitations.

\emph{B.Narwal and A.K.Mohapatra} in \cite{narwal2021survey} presented a systematic survey on security and authentication in Wireless Body Area Networks (WBAN) to cover the main research elements. In particular, an in-depth classification of protection mechanisms in WBAN is provided along with a thorough analysis of security basics, threats to security, the intruder and their attack strategies, and current mitigation that include ML. 

The survey presented by \emph{A.Saxena and S.Mittal} in \cite{saxena2022internet} explained the IoMT ecosystem, reviewed the security requirement and vulnerability of IoMT systems, and presented recent security applications to protect this ecosystem, including blockchain, physically unclonable function and AI/ML.

Among the review papers conducted on security in IoMT, only a few survey papers have reported using ML as a method for security purposes in the literature. They describe it briefly and not in a holistic manner. However, a work similar to the current study conducted by  \emph{S.S.Hameed, et al.} in \cite{hameed2021systematic} discussed security and privacy in IoMT. They presented the different solutions based on ML proposed in the literature to solve the security challenge of IoMT, giving their advantage, disadvantage, approach, tools, and datasets used.  However, the authors focused on network and device-level security and did not discuss the security of electronic medical data when it resides at the medical server level.  
This paper extends the previous work and discusses the various solutions used for anomaly detection based on ML at the medical device level, at transit, and during the resets by identifying the benefits, drawbacks, and datasets used. Reviewing the solutions at the three levels allow for discussing the ML method used to ensure the security and privacy of the IoMT globally.

\begin{table}[width=\linewidth,cols=4,pos=H]
\caption{Existing Surveys}\label{tbl1}
\begin{tabular*}{\linewidth}{ m{1cm} m{1cm} m{3cm} m{10cm} }
\toprule
Year & paper & Discussed topic(s) & Main differences\\
\midrule
2017 & \cite{rathore2017review} & Security of wireless medical devices & This survey provides the security and proposed solutions for the three layers composing the IoMT system\\
\midrule
2019 & \cite{hussain2019authentication} & Authentication in the wireless body area network & This survey presents the architecture, attacks on each layer that compose the IoMT, the environment, and the attackers’ motivations. Then, the different ML-based solutions proposed for the layers forming the IoMT system are discussed \\
\midrule
2019 & \cite{wazid2019iomt} & Detection and prevention of malware in IoMT & This survey mentioned the attacks that can occur in each layer that composes IoMT, the types and environments of attacks, and the attackers' motivations, not only the malware.  The IDS-based ML solutions proposed for the three layers that constitute the IoMT system are discussed in this survey and do not only cover the security of the communications \\
\midrule
2020 & \cite{newaz2021survey} & Security and privacy in the healthcare system & The IDS-based ML solutions for the three layers of the IoMT system are provided \\
\midrule
2021 & \cite{narwal2021survey}  & Authentication in the wireless body area network & The different attacks occurring at the three layers composing the IoMT system are listed, and the solutions that use an IDS based on ML and proposed for the three layers of the IoMT system are discussed \\
\midrule
2022 & \cite{saxena2022internet}  & Review the use of new technologies to improve IoMT ecosystem security & The security requirements and threats for the three layers that compose the IoMT systems are provided. Then, the IDS solutions based on ML for each layer that forms  the IoMT systems are reviewed, which allows identifying the limitations and challenges at each layer of the IoMT systems \\
\midrule
2021 & \cite{hameed2021systematic} & The role of ML in solving the security and privacy issue in IoMT systems & The different attacks that can impact the three layers of the IoMT and the IDS based on ML solutions proposed for layer three of the IoMT system are presented \\

\bottomrule
\end{tabular*}
\end{table}

\section{ARCHITECTURE OF IoMT} \label{ARCHITECTURE OF IOMT}
Many architectures were presented in the literature. Some research proposes architectures composed of three layers \cite{rahmani2018exploiting, irfan2018internet}. Other research proposes using an architecture with more than three layers \cite{khan2012future}. Different technologies are proposed to manage medical data, such as fog/cloud computing \cite{kumar2021ensemble}, Software-Defined Networking (SDN) \cite{khan2021hybrid}, or Blockchain \cite{chakraborty2019secure}. This survey paper assumes that a three-layered architecture is suitable for logically dividing IoMT architecture. These layers are data acquisition, personal server, and medical server, and they are explained as follows and illustrated in figure \ref{FIG:2}.

\subsection{Data Acquisition Layer}
Sensor devices serve as a bridge between the human body and the digital world. There are four types of devices \cite{irfan2018internet, khan2012future}:

\begin{itemize} 
\item The implemented devices: these devices are placed inside the human body, e.g., DBI.
\item The wearable devices: these devices are on the human body e.g., Smartwatch, Pulse Generator (PG), or Electroencephalogram (EEG).
\item The ambient devices: these devices allow capturing data from the environment around the patient, e.g., room temperature sensors.
\item The stationary devices: these sensors are located in the hospital, e.g., imaging devices.  
\end{itemize}

These devices are equipped with physiological sensors and low-power computing, connectivity, and storage modules, which are used to gather biomedical and context signals \cite{moosavi2015sea}. The data collected are used to manage the treatment and diagnosis of patients' health conditions. These medical devices have other constraints at the internal and communication levels. Internally, the medical implants inside the human body can be rejected by the patient's immune system, resulting in inflammation and pain. The battery of these medical devices is limited and must be changed after a number of years. Traditional security methods such as cryptography can quickly shorten the implant's lifespan, which requires surgery to replace the battery and can be dangerous for the patient. Because of medical equipment's limited memory space, it is impossible to keep track of the exchanges made via log files. Regarding communication, medical equipment can only support short-range communication due to energy limitations \cite{rathore2017review}. Due to scale, computing capacities, and energy constraints, most wearable devices can only preprocess the sensed data. Alternatively, the embedded low-power computing modules compress the sensed data before sending it to personal devices (i.e., smartphones or desktops) through low and ultra-low-power wireless communication like Near Field Communication (NFC) and Bluetooth Low Energy (BLE) \cite{zhang2015security}.

\subsection{Personal Server Layer}
Medical equipment can send physiological data to personal servers like smartphones and laptops or standard devices like gateways \cite{sun2019security}. Personal servers process and save patient data remotely until it is sent to centralized medical servers. The medical data received at the personal server are saved and processed; for example, by adding contextual information such as place and time to detect unusual behavior, these data also can be encrypted or compressed. Then the processed data are sent in a medical standard format like Health Level-7 to the remote medical server using long-range wireless like Wireless Fidelity (Wi-Fi), Global System for Mobile Communications (GSM), or wired communication \cite{hussain2019authentication}. This layer is intended to support heterogeneous communication and node mobility and permit the resending of medical data when the network link to the medical servers is interrupted \cite{arya2020data, wazid2019iomt}.

\subsection{Medical Server Layer}
It consists of a high-performance data center that allows for centralized patient control, complex and long-term behavior analysis, and the correlation of patient data. It also includes a cloud server that makes intelligent decisions. It is used for data aggregation and provides extra patient medical data storage. The doctor, patients, and the pharmacy department (for summary or billing purposes) can access these data. Patients may use an online interface or smartphone to display their past and current health records/bills. Data from various sources are incorporated into EHR, Electronic Medical Records (EMR), or prescription websites. As a result, doctors and patients can access the information whenever needed. It provides a notification service for any patient who uploads or receives health data \cite{hathaliya2020exhaustive}.

\begin{figure*}[!ht]
	\centering
\includegraphics[width=\textwidth]{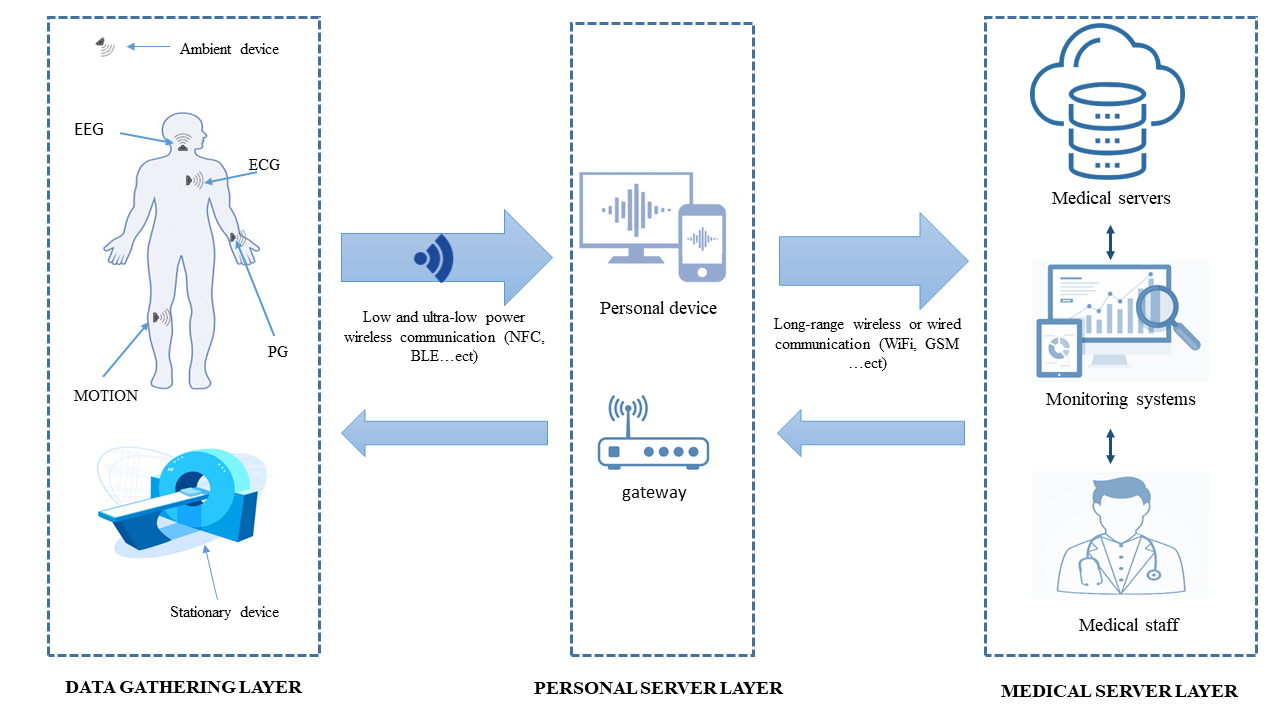}
	\caption{Internet-of-Medical-Things (IoMT) Architecture} \label{FIG:2}
\end{figure*}

\section{SECURITY REQUIREMENTS OF IoMT} \label{SECURITY REQUIREMENTS OF IOMT}
\begin{figure*}[!ht]
	\centering
	\includegraphics[scale=0.7]{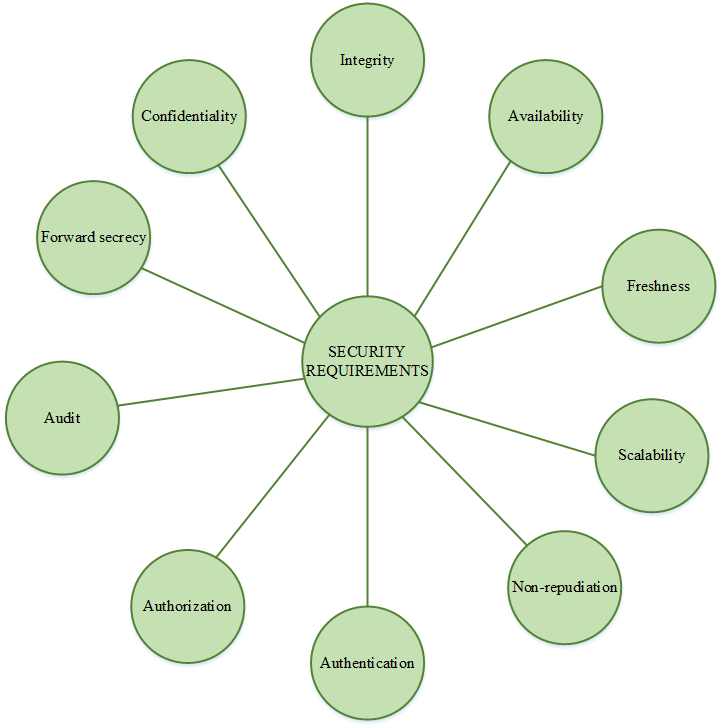}
	\caption{Security requirements in IoMT}
	\label{FIG:3}
\end{figure*}

IoMT uses wireless communication and the internet to transmit data collected from the human body to the medical server; therefore, the data in the different layers of the IoMT system, as shown in figure \ref{FIG:2}, are exposed to cyberattacks, which can impact the privacy of the patients and put their lives in danger. Security requirements must be met to prevent, detect and respond to these attacks in real 
time. This section presents the main security requirements of IoMT (Illustrated in figure \ref{FIG:3}):

\subsection{Confidentiality}
This requirement ensures that the information regarding the patient's health condition or treatment and information that identifies them are not accessible by unauthorized third parties during data storage and data communication; this ensures that the patient's privacy is protected from the disclosure of sensitive information to persons with malicious intent who may cause considerable harm to the patient \cite{wazid2019iomt, alkeem2017new}. It is possible to imagine a scenario in which an adversary, who has access to the medical history of a famous person, discloses it in public to impact his/her image.

\subsection{Integrity}
The data integrity provision for IoMT healthcare systems aims to ensure that the data arriving at the intended destination was not corrupted during wireless transmission. For instance, even a minor change to the medication or patient test results may have catastrophic implications for the patient's life. Preserving the integrity of information ensures that someone other than the person involved (i.e., doctors or nurses) does not change the medical data and, as a result, prohibits giving incorrect treatment \cite{sun2019security}.

\subsection{Availability}
Despite the implementation of healthcare, clinical records of patients must be available to the doctor at all times, anywhere, without any interruption. In addition, it is crucial to respond immediately to the emergency so that the doctor can give the patient treatment or precautions. Switching from the attacked node to another node in the network may be an alternative, and this redundancy can be allowed by the network and system design \cite{sun2019security, arya2020data}.

\subsection{Freshness}
This layer ensures that the health information is recent and that an attacker cannot replay old medical data. Two kinds of freshness are present; weak and strong freshness. Weak freshness gives the partial ordering of the health freshness message, and no delay information is provided. In contrast, strong freshness gives the complete ordering of the medical data and allows the calculation of delay \cite{hussain2019authentication, roy2020security}. For illustration, a physician must be aware of the present patient's vital signs, such as their oxygen saturation, in order for the physician to make a correct diagnosis of the patient's health condition and provide the appropriate treatment.

\subsection{Scalability}
Scalability is the ability of the IoMT network to function properly; as the number of devices that compose the IoMT network turn to be larger, insufficient scalability could cause security flaws. Therefore, it is essential to manage overhead computing and storage, especially in an emergency where response time is vital for the patient \cite{pirbhulal2019joint, kompara2018survey}.

\subsection{Non-repudiation}
Non-repudiation ensures that any entity involved in the healthcare application cannot deny the sending and receiving health-related patient information \cite{narwal2021survey}.

\subsection{Authentication}
There are two kinds of authentication in healthcare systems: data and persons. Data authentication is the process by which the initial data source is confirmed. Person authentication in communications between patients and related servers should be checked through accurate identity authentication. Therefore, before they communicate or exchange any details, all parties need to know each other. Before doing some form of sharing, the healthcare system has to recognize each participant to ensure that the user is authorized to receive the stored data. It is, therefore, essential to know the privilege level given to the user to know the kind of data he/she may access \cite{sun2019security, newaz2021survey}. 

\subsection{Authorization}
Authorization is an access control used to specify permission levels for users (patients, physicians, or nurses) to enter the database of medical data.  The healthcare organization must approve the patient and define the type of data that a single user can use \cite{alkeem2017new, narwal2021survey}.

\subsection{Audit}
The audit is the inspection of changes in the system and access to the patient's medical data via the verification of log files, which are historical records of the hardware and software operating status. The audit allows the detection of abnormal activities and possible breaches.  However, the management and exploitation of this type of information are delicate in practice due to the large quantity and heterogeneity of logs generated by the various medical and network equipment.

\subsection{Forward Secrecy}
\begin{itemize}
    \item Backward secrecy: Medical sensors who enter a network after a certain amount of time must not decode messages received before entering them \cite{kompara2018survey}. 
    \item Forward secrecy: Medical sensors that have left the network are unable to read messages received after their exit \cite{kompara2018survey}.
\end{itemize}

\section{IoMT SECURITY THREAT} \label{IOMT SECURITY THREAT}
The integration of wireless communication in the IoMT system and external equipment to control and upgrade sensors makes the IoMT vulnerable to different attacks. Moreover, two types of architecture are proposed in the literature for the IoMT: the single-hope and the multi-hope. For the single-hope, the sensors perform only data collection and transfer; however, this architecture suffers from the vulnerability of a single point of failure, which occurs when one equipment of the personal server layer fails, the whole IoMT system is compromised. The other type of architecture proposed is the multi-hope, here the sensors, in addition to collecting and transmitting data, also provide data routing, which allows node mobility and maintains low energy consumption during data transfer, like Codeblue \cite{malan2004codeblue} and MIDiSN \cite{ko2010medisn}. Therefore, relevant vulnerabilities of the wireless sensors network that concern routing may apply to this type of architecture \cite{segovia2013analysis}.

Attacks can indirectly target devices in the personal server layer to reach patient data. The lack of data storage security in the personal and medical servers and the insecure transmission of data between these different devices can cause various security issues  \cite{kintzlinger2019keep}.

It is imperative to keep these different attacks in mind when designing a secure architecture for IoMT. The different threats relateds to each level of IoMT are summarized in table \ref{tbl2}. The most common types of network and system attacks are listed below:

\subsection{Attack at Data Collection Level}
During data perception and delivery over a wireless channel and the ability of medical equipment to be remotely configured by external devices led the medical sensors vulnerable to different attacks. Here are the potential security threats that could take place in the following manner:

\subsubsection{Data Modification}
An attacker modifies the medical data of patients \cite{bangash2017security} to manipulate the medical diagnosis and lead to the administration of inappropriate medication, which can be dangerous for the patients.

\subsubsection{Drain the Battery}
Process that leads to additional battery consumption of a medical device, which leads to battery failure \cite{rathore2017review}.

\subsubsection{Modify the Software of the Device}
An attacker can try to modify the software of the medical equipment by introducing a virus that modifies its behavior to conduct malicious actions \cite{hei2014patient}, such as instructing the pacemaker to send an electrical charge to the patient's heart.

\subsubsection{Jamming Attack}
This attack interferes with the radio frequencies that the network's nodes use. A jamming source can interrupt the whole network or part of it. It may be deliberately or unintentionally made \cite{segovia2013analysis, bangash2017security}.

\subsubsection{Node Tampering}
In the case where an attacker has physical access to the IoMT nodes, they can gain access to the sensitive information on the node, for example, retrieve the encryption keys and then use them to decrypt the communications in transit. The attacker can even modify the node or replace it with a different node that the attacker controls \cite{arya2020data, doss2015future}.

\subsubsection{Data Collision Attack}
 When two nodes of IoMT system transmit on the same frequency simultaneously, a collision may occur, which implies a modification of the transmitted data. Therefore, the receiving node of the packet will reject it due to the checksum mismatch \cite{segovia2013analysis}.
 
 \subsubsection{Exhaustion}
 The collision implies that the packet lost is retransmitted; this retransmission consumes energy at the medical equipment level. The intruder could exploit the repeated collision to create resource exhaustion \cite{segovia2013analysis, roy2020security}.

\subsubsection{Unfairness in Allocation}
Collision and exhaustion attacks can be used by the intruder to create unfairness in the network, Therefore, when incoming patient data packets enter the application's processing device, they are either missed or produce multiple errors \cite{rani2017survey}.

\subsection{Attack at the Transmission Level}
There is a high risk of threat at the transmission level because wireless communication allows an attacker to intercept, modify, or block the messages sent and exchange valuable information related to the patient's condition. Some risks include:

\subsubsection{Eavesdropping}
Due to the communication over wireless networks, all traffic is vulnerable to detection and eavesdropping by attackers. These threats may result in the loss of personal information such as physiological data and may obtain information about the medical device, such as the type of medical equipment associated with the patient. They can also contribute to other kinds of attacks. There are two types of eavesdropping:
\begin{itemize}
\item Passive eavesdropping: By listening to the network's message transmission, a hacker will intercept the content \cite{xu2016data, Sanei2020}.
\item Active eavesdropping: By pretending to be a friendly entity and sending requests to emitters, a hacker actively obtains information \cite{xu2016data, Sanei2020}.
\end{itemize}

\subsubsection{Man In the Middle Attack}
This attack happens when an intruder gets in between a patient and a server's communications and sniffs the data. They can capture all messages received between the two parties and insert new ones \cite{Sanei2020}.

\subsubsection{Scrambling Attacks}
This attack keeps the radio frequency channel for wireless communication occupied for a brief period of time; this causes patients' personal devices to be interrupted to block data transfer, resulting in vulnerabilities of losing the network availability \cite{arya2020data, habib2015security}.

\subsubsection{Signaling Attacks}
In this threat, the attacker will try to disrupt the signaling operation that takes place before the establishment of a communication between two entities and involves key management, authentication, link creation, and registration by sending an additional signal, which will put a heavy load on the base station and thus interrupt its service. As a result, health data cannot be forwarded to their destination \cite{maheswar2019body}.

\subsubsection{Message Modification Attack}
By capturing patients’ wireless channels, the attacker can extract health data from patients, which can then be partially or fully altered before being delivered back to the initial recipient \cite{kumar2012security, pathania2014security}.

\subsubsection{Data Interception Attack}
An attacker can capture a patient's health data when it is transmitted between two connected healthcare devices over a local area network \cite{habib2015security}.

\subsubsection{Wormhole Attack}
In this attack, a duplicate of a patient's data packet from one location is replayed at a different location with no modifications to the data. This attack usually requires two rogue nodes relaying packets over an out-of-bound channel that is only accessible by the attacker \cite{segovia2013analysis, maheswar2019body}.

\subsubsection{Denial of Service Attack (DoS)}
In this attack, the intruder tries to interrupt, shut down, or stop a service from a system, machine, or a network. A distributed DoS (DDoS) attack, on the other hand, is a kind of DoS attack that originates from several distributed sources \cite{kintzlinger2019keep, Sanei2020}.

\subsubsection{Hello Flood Attack}
The attacker must persuade all nodes to choose them as the patient's medical data routing node. A malicious node accomplishes this by sending a HELLO packet over a strong radio transmission to the network. When a node receives a message like this, it will presume that the source is within a normal radio range. Such an assumption could be wrong, since the machine of the intruder with high transmission capability could mislead the target into assuming that it is his neighbor \cite{maheswar2019body, stavroulakis2010handbook, hamid2006routing}.

\subsubsection{Sinkhole Attack}
The sinkhole attack occurs when a malicious node attempts to attract all data packets in IoMT system by pretending that it is the most appropriate routing algorithm. This attack is performed to prevent data packets from reaching their destination \cite{stavroulakis2010handbook, tumrongwittayapak2009detecting}.

\subsubsection{Replaying Attack}
In a replay attack, an attacker listens to communications received between valid entities in IoMT system, intercepts them, and then sends them again to an initial recipient to alter the overall result \cite{narwal2021survey, mana2009sekeban}.

\subsubsection{Homing Attack}
A scan is performed in the continuous data traffic to find the key manager or the cluster head that can stop the whole network of IoMT system \cite{stavros2002advances}.

\subsubsection{Data Flooding Attack}
This attack occurs when an attacker sends many connection requests to a target node in the IoMT network. As a result, the server exhausts its capabilities and cannot establish any other connections, even legitimate ones \cite{doss2015future, siva2019security}.

\subsubsection{Desynchronization Attack}
The intruder repeatedly resends an incomplete message to one or both nodes that participate in communication within IoMT system, which requests the retransmission of the missing data. Consequently, valuable information is prevented from being exchanged between the endpoint \cite{gupta2020handbook}.

\subsubsection{Spoofing Attack}
The attacker manipulates the routing information to compromise the network of IoMT system \cite{masdari2016comprehensive}.

\subsubsection{Selective Forwarding Attack}
In this attack, the attacker will target a node in the IoMT network to compromise it and make it malicious, transmit some messages and block the rest. The number of messages lost will be more important if the compromised node is closer to the base station; therefore, many vital data will be wasted  \cite{arya2020data, niksaz2015wireless}.

\subsubsection{Sybil Attacks}
A Sybil attack occurs when a given node in the IoMT network claims several identities to act (modify) geographical routing protocols \cite{newsome2004sybil}.

\subsubsection{Path DoS Attack}
In this form of attack, the intruder produces a significant volume of traffic to the base station \cite{gupta2020handbook}.

\subsubsection{Overwhelming Sensors}
In this threat, the intruder caused the network capacity dilapidation and node resource exhaustion by overwhelming the network node with sensor stimuli, causing the network to send a large amount of traffic to the base station \cite{roy2020security}.

\subsubsection{Reprogramming Attacks}
TinyOS' Deluge network-programming framework, for example, enables nodes in deployed networks to be remotely reprogrammed. The majority of these systems, like Deluge, are meant for use in a protected environment. A hacker will hijack the reprogramming mechanism and take possession of a vast network area if it is not secured \cite{roy2020security}.

\subsection{Attack at Storage Level}
All information related to the patient’s health condition, treatment, and identity are stored at this level, which constitutes a valuable target for adversaries to access these data. Some of the possible attacks that can occur are:

\subsubsection{Patient’s Data Inference}
Intruders try to recover medical records such as information about patient’s health, maladies, and medications by combining information that the attacker is authorized to access with other pertinent information \cite{maheswar2019body}.

\subsubsection{Unauthorized Access}
If the patient data are not protected, the attacker will attempt to access health data to conduct a malicious action such as damage it or retrieve it; therefore, it is important to secure the data against unauthorized access \cite{ramli2010privacy}.

\subsubsection{Malware Attack}
Malware (short for “malicious software”) is a code or program usually spread over the network. It extracts, infects, or executes other malicious operations directed by the attacker. The types of malware include viruses, spyware, Keylogger, worm, rootkit, ransomware, and Trojan horses \cite{wazid2019iomt}.

\subsubsection{Social Engineering Attacks}
Phishing, spear-phishing, baiting, and quid quo pro are examples of techniques to gain sensitive information from victims. These techniques are used to dupe the user into supplying the attacker with sensitive patient information that the user assumes to be supplied to someone or something else \cite{habib2015security, partala2013security}.

\subsubsection{Location Threats}
Most medical devices are equipped with a location component that allows the caregivers to have a quick response in an emergency. If this type of information is not well protected, adversaries can access it and directly invade a person’s privacy \cite{kumar2012security, curtis2008smart, redondi2010laura}.

\subsubsection{Alert Attack}
Some medical devices are equipped with an alert system that notifies the medical staff of an abnormality concerning the patient's health or a device's malfunction, such as a necessary battery change. However, this feature can be abused by an attacker to create false alarms; therefore, the wrong treatment can be prescribed, the patient can make unnecessary hospital visits, or genuine alerts can be ignored, and the patient can even disable this notification functionality which has the effect of missing important notifications \cite{kintzlinger2019keep}.

\subsubsection{Extortion/Blackmail}
This type of attack consists of blocking access to medical data by an attacker and then asking for a ransom to unlock these data. It can also involve stealing influential people's personal medical history, such as a politician or a star, and then threatening to disclose it publicly if the person concerned does not pay the amount of money requested \cite{mcglade2019ml}.

\begin{table}[width=\linewidth,cols=3,pos=ht]
\caption{Threat attack at each level of IoMT}

\begin{tabular*}{\tblwidth}{m{3cm} m{0.1cm} m{10cm} }
\toprule
Data Collection Level & & Data modification\\
 & & Drain the battery\\
 & & Modify the software of the device\\
 & & Jamming attack\\
 & & Node tampering\\
 & & Data collision attack\\
 & & Exhaustion\\
 & & Unfairness in allocation\\
\midrule
Transmission Level & & Eavesdropping\\
 & & Man In the middle attack\\
 & & Scrambling attacks\\
 & & Signaling attacks\\
 & & Message modification attack\\
 & & Data interception attack\\
 & & Wormhole attack\\
 & & Denial of service attack\\
 & & Hello flood attack\\
 & & Sinkhole attack\\
 & & Replaying attack\\
 & & Homing attack\\
 & & Data flooding attack\\
 & & Desynchronization attack\\
 & & Spoofing attack\\
 & & Selective forwarding attack\\
 & & Sybil attacks\\
 & & Path dos attack\\
 & & Overwhelming sensors\\
 & & Reprogramming attacks\\
 \midrule
 Storage Level & & Patient’s data inference\\
 & & Unauthorized access\\
 & & Malware attack\\
 & & Social engineering attacks\\
 & & Location threats\\
 & & Alert attack\\
 & & Extortion/Blackmail\\
\bottomrule
\end{tabular*}
\label{tbl2}
\end{table}

\subsection{Type of Attack}
Attack strategies are continually changing. However, they can be divided into passive and active attacks. 

\subsubsection{Passive Attack}
In a passive attack, the adversaries will only listen to the traffic and thus will have the possibility to read messages exchanged between the wearable device and the remote system. By simply accessing the content of the messages, a passive attacker will directly affect the confidentiality of communication. They will have access to sensitive information such as the model, the serial number of the medical device, and capture telemetry data. They can also capture the patient's private data such as health record, name, age, and conditions. In all these cases, the result is a severe violation of the patient's privacy \cite{kintzlinger2019keep}.

\subsubsection{Active Attack}
The attacker will intercept network messages and give instructions to the wearable device, alter messages transmitted before they reach the remote system, or prevent them from reaching their intended destination. A successful intruder has a wide range of objectives. They might, for example, indiscriminately request information from the medical device to deplete its energy. They may even try to change the device's settings, bypass treatments, or even put the patient in a state of shock \cite{kintzlinger2019keep}.

\subsection{Attack Environment}
Threats to the system can be classified according to the adversary's position, i.e., internal attack and external attack.

\subsubsection{Internal Attackers}
Internal attackers require that the attacker is close to the vulnerable device or nearby and has some rights to enter the network infrastructure. They may be a legitimate user, like a nurse who accesses a celebrity patient's medical data without justification. The attacker near the medical equipment can then cause physical damage or collect some information and use them to launch remote attacks later \cite{narwal2021survey, alsubaei2017security}.

\subsubsection{External Attackers}
In this case, the attacker does not need to be close to the medical device and does not have administrative access to the system. Instead, they will try to exploit bugs or vulnerabilities of the system remotely \cite{narwal2021survey, alsubaei2017security}.

\subsection{Attackers Motivations} \label{Attackers_motivations}
This section discusses various attackers' motivations for targeting IoMT systems \cite{hassija2021security}.

\subsubsection{Physical Injury}
The compromise of medical equipment poses a severe threat, as it could be used to cause harm to patients. Malicious organizations may target patients for political or criminal reasons, and in some cases, even by terrorist groups. Such attacks can be powerful tools for criminal practices, such as extortion or coercion. For example, during his tenure, former US Vice President Dick Cheney had the wireless communication functionality of his pacemaker deactivated as a precautionary measure against potential hacking attempts \cite{DanaFord2013}.

\subsubsection{Financial Profits}
Economic and financial profits are important motivators for attackers or rivals of Implementable Medical Device (IMD) vendors to conduct such threats. The ability of an attacker to access medical data may be used to sell it or blackmail the patient.

\subsubsection{Privacy Violation}
 Medical equipment collects vital information about a patient's body based on various criteria. This information may be necessary for the patient's diagnosis, care, and operational or surgical procedures. Medical data divulge information about the patient's behavior. Analysis of the data obtained from a pacemaker, for example, will reveal the patient's physical activity history. Such data can be used to distinguish general and unique patterns in the well-being of individuals/groups if collected in a large enough sample across different device types and marks. This type of information can lead to unauthorized and unethical use of sensitive data.

\subsubsection{Tracking}
Messages from the Medical equipment are sent wirelessly to the system controller. These exchanges typically provide health data and position information about the patients. Attackers can intercept these communications to track or locate a patient.

\section{INTRUSION DETECTION SYSTEM} \label{INTRUSION DETECTION SYSTEM}
This section explains the IDS in IoMT systems according to \cite{bace2001nist}. An intrusion is an attempt to compromise the availability, integrity, confidentiality, or defeat the security mechanisms of an end device or network. The IoMT represents information system that handles sensitive and private data related to the health data of patients, which constitutes a valuable target for an attacker to perform an intrusion. This intrusion can be performed by a remote attacker using the Internet or by a legitimate internal user who abuses their privileges, such as members of the medical team who are motivated by curiosity to access private data or an error in the handling of medical data, which could have severe repercussions to patients. An IDS is a hardware or software product that automates surveillance and analysis of events that occur in an end device like IMD or network to detect signs of intrusion.
    
An IDS consists of three parts: information source, analysis, and response. An IDS can use several information sources to perform a pre-configured analysis on them. When an attack is detected, the IDS generates a response that can be passive or active. A passive response implies issuing a notification. An active response does an action such as interrupting communication. The purpose of an IDS is not to find out who conducted the attack. Instead, it is to interrupt it because the attacker's identity may be hidden, making it difficult to Recognize. 

An IDS differs from other security mechanisms, such as a firewall or an antivirus, in that IDS monitors network traffic and analyzes events to identify potential threats, while firewalls and antivirus software primarily focus on blocking or detecting known threats. However, each security mechanism has its advantages and disadvantages, and their combination can provide in-depth security capable of protecting information systems from various security threats.

Several types of IDSs are classified based on their monitoring approach, information source, type of analysis performed, and response time. There are network-based IDS for the monitoring approach and host-based IDS. Network-based IDS monitors network packets by listing network segments. Host-based IDS is designed to monitor events and activities on individual end devices, including the operating system, audit trails, system logs, and other data sources, such as medical records. There is also an application-based IDS, a subdomain of the host-based IDS, which monitors the application transaction log file generated by applications to perform attack detection.

There are two types of analysis methods used by IDS: misuse detection and anomaly detection. Misuse detection identifies attacks by matching the analyzed event to predefined patterns, which ensures high accuracy. However, this method cannot detect new or zero-day attacks, and new attack signatures must be added to the predefined templates continuously. On the other hand, anomaly detection identifies suspicious behavior by learning the normal operating pattern of the system using historical data. Any deviation from this normal behavior is flagged as potentially malicious. This method has the advantage of detecting new vulnerabilities and zero-day attacks. However, it can generate many false positives and requires a large training set to build the normal profile. 
 
The response time of IDS can be real time or interval-based.

There are several possible architectures for IDS:  centralized, fully distributed, and partially distributed. In a centralized architecture, monitoring, detection, and reporting are performed in a central node. In a fully distributed architecture, the response to an intrusion is carried out in the part of the network where monitoring is conducted. In a partially distributed architecture, the reporting is executed hierarchically.

\section{MACHINE LEARNING OVERVIEW} \label{MACHINE LEARNING OVERVIEW}
ML is a subfield of AI that gives a machine the ability to learn from data without explicitly programming it \cite{samuel2000some}.  ML has proven efficient for problems requiring a long list of rules and complex problems where traditional approaches are inefficient. Also, ML shows a great capacity for adaptation to new data, especially in a changing and evolving environment such as the IoT system. ML has also shown a great ability to obtain insights from large volumes of data \cite{geron2017hands}.

These ML capacities can be used to enforce security in the IoMT or at least improve it. Some studies demonstrate that using ML in IDS effectively detects zero-day attacks and new vulnerabilities, while IDS that rely on rules and signatures can only detect known attacks. 

ML can also be used to learn the behavior of a person or an object by using the data generated to create a so-called normal profile, so any behavior that deviates from the normal profile is considered abnormal and consequently is detected, this is what is done in the field of anomaly detection.

Three types of ML can be used to solve security problems in the IoMT: supervised learning, unsupervised learning, and semi-supervised learning.

For supervised learning, the training data are labeled. The relationship between inputs and their appropriate output
is captured. For this, there is a need to train a model with
labeled inputs, which are used to predict or classify new data
\cite{franklin2005elements}. There are two types of supervised algorithms, which are regression and classification. The regression predicts the next continuous value based on the previous ones. On the other hand, classification predicts discrete variables and separates the data into different categories. Examples of the techniques that can be used for regression and classification are Support Vector Machine (SVM), Decision Tree (DT), Random Forest (RF), Naive Bayes (NB), and artificial neural network (ANN).

Unsupervised learning involves training ML models on unlabeled data, as manually labeling data can be challenging and time-consuming. Instead, unsupervised learning algorithms analyze the data and identify patterns, relationships, or groupings among the data points. Some popular unsupervised learning techniques for clustering data include K-means, k-Nearest Neighbor (KNN), and Self-Organizing Map (SOM). Unsupervised learning can effectively divide large datasets into meaningful clusters to aid further analysis or decision-making \cite{geron2017hands}. 

While obtaining unlabeled data is often straightforward, leveraging its potential can be a significant challenge. Semi-supervised learning presents a solution by incorporating labeled and unlabeled data to build ML models that can classify data with improved accuracy. Compared to traditional supervised learning, semi-supervised learning reduces the need for manual labeling while producing models with greater accuracy. As a result, this approach has a special place in theoretical research and practical applications \cite{pise2008survey}.

DL is a subdomain of ML, and it is inspired by the functioning of the human brain to process the signal.  DL enables computational models with several layers of processing to learn data representation with several levels of abstraction \cite{lecun2015deep}. DL differs from classical ML in its ability to capture the relevant feature previously done manually and required human intervention. Another advantage of DL compared to traditional ML is its performance on a large dataset. Therefore, this method is perfectly adapted to IoT systems and their applications, such as IoMT, which generates a huge volume of medical data. There are three types of DL: supervised example Convolutional Neural Network (CNN), unsupervised example the Deep Autoencoder, and finally, the combination of these two types, which is the hybrid DL, example ensemble of learning networks.

\section{REVIEW OF MACHINE LEARNING APPLICATIONS FOR IoMT SECURITY} \label{REVIEW OF MACHINE LEARNING APPLICATIONS FOR IOMT SECURITY}
Health data are sensitive and can attract the attention of attackers for various reasons mentioned in section \ref{Attackers_motivations}. It is essential to protect medical data from threats by adopting a solution that ensures the security of this data during collection, transfer, storage, and processing. Protecting patients’ privacy from unauthorized access is also necessary to avoid disclosing sensitive information to malicious entities. In addition, security solutions must consider the computational, storage, and energy limitations of medical devices and the heterogeneity, dynamics, and quantity of medical data generated within the IoMT system. In light of these constraints, ML has the potential to provide a solution for intrusion detection.

This section reviews the papers that have proposed a security solution based on ML for the different IoMT layers, specifying their objectives, the method employed, the dataset used, and the results obtained. Tables \ref{tbl3}, \ref{tbl4}, and \ref{tbl5} summarize the papers explored in this review, mentioning the methods, applications, advantages, disadvantages, and dataset. 

\subsection{ML for Data Collection Level Security}
Sensors associated with patients allow constant monitoring of their health status and automatic medication for people suffering from chronic diseases. This technology, despite its advantages, must be protected against security threats due to the wireless transmission of medical data to a personal server and the possibility of configuring the medical equipment via a programmer device, which increases the surface of the attack. These security threats can be either intentionally caused by hackers or unintentionally caused by the environment due to interference. Patients need a security solution that detects such threats, differentiates them from emergency cases and avoids lethal doses of medication.

Among the different medical devices that researchers have investigated to provide a security solution, there are cardiac implants that provide electrical impulses to stimulate the cardiac muscles in case of abnormalities in the heart rhythm. A well-placed attacker can cause an electrical impulse that can be fatal for the patient. To avoid this kind of situation, the researchers (Kintzlinger et al., 2020) in \cite{kintzlinger2020cardiwall} proposed a system that detects and prevents cyberattacks against ICD. It is implemented at the level of the programmer's device. The Cardiwell is a decision aid for doctors that, in case of detection of an anomaly, generates an alert with the necessary details that allow the doctors to decide then if they pass the programmer commands. The Cardiwell is a multi-layer Security scheme, and it consists of six layers of security. The first five layers use rules and statistics, while the sixth layer uses the method of one class SVM, which is based on ML. To validate their solution, tests were performed on a dataset collected from volunteer patients from different hospitals and clinics over four years. This dataset consists only of benign programmer commands sent from a programmer device to an ICD and which represent a total of 775 samples, while experts generate malicious programmer commands and represent a total of 28 samples. After performing the different tests, the best results obtained are True Positive Rate (TPR) = 0.914, False Positive Rate (FPR) = 0.101, and Area Under the Curve (AUC) = 0.947. However, layers three to six are inefficient and do not participate in improving the results, which according to the authors, is due to the poor datasets that need to be enriched.

The study by (Khan et al., 2017), as described in \cite{khan2017continuous}, presents a novel approach for detecting abnormalities in Electrocardiogram (ECG) data. The authors proposed a centralized solution and chose a simplified Markov model-based detection mechanism due to the changing nature of ECG data over time. The proposed approach involves extracting attributes from the ECG data, followed by reducing the dimensionality of the dataset using the discrete wavelet transform. The reduced dataset is then partitioned into sequences, and the probability of each sequence is calculated to determine if any changes have occurred.
If an abnormality is detected, the system associates an abnormal tag with the corresponding data and forwards it to the hospital server for further evaluation by nursing staff. The evaluation process is designed to determine the normality of the received data. To test the efficacy of their approach, the authors utilized a dataset from MIT-PHYSIOBANK \cite{LugovayaT.S2005} and introduced 5\% and 10\% of attacks composed of forgery, unauthorized insertion, and ECG data modification. After conducting the experiments, the authors achieved a high detection rate, reduced training time, and a high TPR.

Other researchers have studied the security of the insulin pump system, which allows the automatic administration of insulin according to the amount of glucose in the patient’s blood. However, an intruder can cause a lethal insulin administration if the device is compromised. In this perspective, the researchers (Hei et al., 2014) in \cite{hei2014patient} addressed the security gap between the Carelink and the insulin pump, two components of the insulin pump system where the attacker can carry out two types of attacks; the first one is the bolus dose where the attacker delivers a large amount of insulin in a short period. The second possible attack is the basal dose, where the attacker delivers an insignificant amount of insulin over a long period, which can threaten the patient's life. 
To address these issues, the authors propose to use a supervised ML algorithm using SVM with a regression method to learn the normal insulin dosage of each patient at different parts of the day. The authors collected insulin pump log files from four patients over six months to generate the dataset. Each log entry is composed of infusion rate, dosage, blood glucose level, patient ID, and time of day for each infusion. After performing different tests, the authors obtained better results with the non-linear SVM than the linear SVM by getting a score of 98\% of success rate in detecting a single overdose attack and a high success rate in chronic overdose attack detection. 
The authors propose deploying the model with an update every 90 days. In the case of anomaly detection, an alert message is raised. The authors have also defined a value of insulin that, if exceeded it, indicates that the patient is in an emergency and, therefore, the system is deactivated. 

In another study made by (Ahmad et al., 2018) in \cite{ahmad2018securing}, The authors proposed to secure the insulin pump against attacks that aim to alter the functioning of its system to deliver a lethal dosage of insulin to the patient. To address this issue, the authors propose using Long Short-Term Memory (LSTM), a DL algorithm, to define a threshold value for the insulin dosage delivered to the patient. If the insulin dosage value exceeds the threshold value, the system calls the patient to perform a gesture by raising the thumb or spanking it down, which is captured by a gesture sensor. The authors propose to use a gesture sensor because they assume that the patient can feel and judge the significant change of insulin in their body through symptoms like nervousness, trembling, and weakness. Therefore, depending on these symptoms and the dose of insulin to be administered, the patient can accept the insulin dose by raising the thumb or refusing it by lowering it. 
To test their solution, the authors used a dataset containing log files generated by the insulin pumps system over the last three months to train and test the LSTM and determine a threshold value.

In a study conducted by (Rathore et al., 2017) in \cite{rathore2017dlrt}, the authors proposed a solution for detecting fake glucose measurements in the entire insulin pump framework, which comprises a continuous glucose monitoring system, a sensor, a transmitter, an insulin pump, a remote control, and a one-touch meter. The authors proposed a solution that uses the Multi-Layer Perceptron (MLP), a DL algorithm, to classify glucose measurements in the blood as genuine or fake.
To evaluate the effectiveness of their solution, the authors used the Pima Indians Diabetes dataset  \cite{VincentSigillito1990} and achieved 93.98\% of accuracy. Furthermore, the authors employed the NB algorithm to test the reliability of their system and obtained a success rate of 90\%. According to the authors, this high success rate is due to the deployment of their model on a chip using Field Programmable Gate Arrays (FPGA), which can be integrated into any equipment used in the insulin pump framework, thereby securing it against erroneous blood glucose measurements.

So far, the methods used to secure the insulin pump system are based on supervised learning methods. However, the study conducted by (Shobana, 2022) in \cite{shobana2022towards} proposed using an IDS based on unsupervised ML to secure the insulin pump injection against four generated attacks: Long Resume, Single Acute Overdose, Underdose, and Chronic Overdose. For this purpose, the authors proposed using a deep autoencoder and a DL algorithm to classify the logs generated within the medical equipment. To test and evaluate their solution, they applied their method to the Diabetes Data Set \cite{UMLRDDS2022}. They measured the model's performance in terms of accuracy, precision, F1-measure, and recall. The results obtained exceed other traditional ML methods and previous work.

In a pioneer work carried out by (Rathore et al., 2019) in \cite{rathore2019novel}, a solution based on ML was proposed to enhance the security of DBI. A DBI comprises a quadripolar electrode implanted in the human brain, an Implementable Pulse Generation (IPG), and a controller that switches on/off the medical device. Some IPGs are customizable to modify voltage levels. The primary function of the DBI is to regulate the Rest Tremor Velocity (RTV) by maintaining its value at zero through electrical charges. Movement disorders and chronic diseases like Parkinson's arise when the RTV deviates from zero. An attacker with access to the DBI can attempt to modify the RTV's value, thereby seriously endangering the patient's life. To address this issue, the authors modified the RTV value to simulate various attacks. They proposed to use LSTM, a DL algorithm, to predict the RTV value at time T. By detecting and classifying any deviations from the predicted RTV value, the proposed solution can prevent potential attacks on the DBI. To create and test the model, the authors used a dataset obtained from Physionet \cite{Goldberger2000}, which contained 173,398 samples with ten features. However, the authors focused primarily on the RTV attribute, considered the most critical attribute since its modification by an adversary could have detrimental effects on the patient. After conducting various tests, the model efficiently classified attack strategies with minimal loss values and training times.

The solutions explored so far have focused on securing a single medical device. This section reviews more general solutions that include multiple medical devices. In this regard, (Hei et al., 2010) in \cite{hei2010defending} suggested adding a security layer for IMD by using patient IMD access pattern and SVM. This solution prevents authentication requests from illegitimate programmers or readers, saves energy, and counters attacks that drain IMD energy resources. If the IMD’s battery power is depleted, surgery is required to replace the battery, which can be life-threatening. The additional security layer implements a ML-based cell phone model using SVM that classifies authentication requests from readers. Depending on the response of the cell phone, three situations are considered: 1) in case the request is classified as benign, then the IMD can perform the authentication with readers, 2) in case the request is classified as malicious, then the IMD put itself on standby to not waste more energy. 3) If the model fails to classify the request, the decision is left to the patient, who can authorize or deny the authentication request. For the emergency, the authors suggest assigning a value to cardiac implants that signify a critical condition for the patient when reached. In this case, they suggest either disabling the classification or using a backdoor accessible via a key shared between the IMD and the authorized persons to guarantee access to the implant. The dataset used to create the model consists of 3000 entries, of which 2500 are used to train the model and 500 to test it. The dataset is composed of five attributes which are the reader action type which determines the type of action that the reader wants to perform on the IMD, a time interval of some reader’s action, location (home, hospital, pharmacy), time, and day (weekend, weekday). After performing the different tests, they obtained an average accuracy of 90\% for the linear SVM and 97\% for the non-linear SVM.

Other researchers (Newaz et al., 2019) in \cite{newaz2019healthguard} presented a secure framework to detect malicious activities in the Smart Healthcare System (SHS). The proposed framework uses different ML: ANN, DT, RF, and KNN, to detect malicious activity in the SHS. The data used to evaluate their framework are collected from eight different databases. The dataset obtained contains 20,000 samples, of which 17,000 represent data from healthy people and people suffering from diseases, and 3,000 represent attacks that simulate three different threats, which are compromised medical devices, DoS, and false data injection. After performing the tests, they obtained 91\% accuracy and 90\% f1-score.

In another work carried out by (Salem et al., 2021) in \cite{salem2020markov}, they proposed a centralized Markov chain-based solution for the detection of anomalies from the data collected by biosensors in WBAN that is composed of sensors and a Local Processing Unit (LPU). In the proposed system, only the measurements captured at the sensors that deviate from the expected values are communicated to the LPU, reducing energy consumption caused by the transmission of routing data. The proposed method is based on a Markov Model (MM) constructed based on the Root Mean Square Error (RMSE) between the forecasted and measured value for complete attributes. The method intends to work with LPU to detect abnormal deviations in the gathered data and reject erroneous or added measurements. After detecting physiology-related changes and removing erroneous or inserted measures, the system alerts the healthcare staff. To test and evaluate their system, the authors used a public dataset containing real data obtained from Physionet \cite{MoodyGB1996}. After performing the different tests, the authors obtained 100\% TPR while maintaining a low False Alarm Rate (FAR) of 5.2\%. They also compared their approach with other existing methods that use the Markov chain for ECG anomaly detection and other supervised ML algorithms: SVM, KNN, J48, and the distance-based method. The system proposed in this paper exceeds the MM-based ECG abnormality detection system by a small margin and outperforms the ML methods regarding accuracy. 

After reviewing the different solutions of security based
on ML for data collection level, most of these researches are focused on the security of cardiac implants \cite{kintzlinger2020cardiwall, khan2017continuous}  and insulin pump injection systems \cite{rathore2017dlrt, hei2014patient, ahmad2018securing}. There is just one study based on the security of DBI \cite{rathore2019novel}. However, other medical implants use wireless communication and are not yet investigated, such as the Gastricelectrical stimulator. This medical device stimulates the smooth muscles of the lower stomach equipment to help control chronic nausea and vomiting associated with Gastroparesis. This equipment uses wireless communication that suffers from a lack of encryption, authentication, validation mechanism, and Hardware/Software error \cite{yaqoob2019security}, making it vulnerable to different attack types, including eavesdropping, information disclosure, tampering, jamming, and resource depletion \cite{rathore2017review}. Therefore, it must be considered when the researchers conceive an IDS for multiple medical devices.

\begin{table}[width=\linewidth,cols=6,pos=H]
\caption{Details of published studies that use ML for security purposes in Data Collection Level}\label{tbl3}
\begin{tabular*}{\linewidth}{ m{1.5cm} | m{1.5cm} | m{2cm} | m{3.2cm} | m{3.7cm} | m{1.5cm} }
\toprule
Ref  & Methods & Application & Advantage & Limitations & Dataset\\
\midrule
(Kintzlinger et al., 2020) \cite{kintzlinger2020cardiwall} 
& 
Rule, statistical and One class SVM 
& 
Detection of cyber-attack against ICD
& 
 - High TPR
\newline - high AUC
\newline - Low FPR
\newline - Real time
\newline - There is no extra consummation of energy and configuration to make in ICD
\newline - Anomaly detection can be done even from legitimate equipment
\newline - Emergency considered
& 
 - The dataset is unbalanced
\newline - Learns from only benign data
\newline - Layer III-V is not needed
\newline - There is no calculation of overhead
\newline - The proposed IDS is for a single type of medical equipment
\newline - The solution is deployed on external equipment that needs protection
&
 - Self-created clinical Data
\newline - Not available
\\
\midrule
(Khan et al., 2017) \cite{khan2017continuous} 
& 
Discrete Wavelet Transform and Simplified Markov model 
& 
Detection of abnormality in ECG data
& 
 - High detection rate
\newline - Real time
\newline - High TPR
 &
 - Calculation of overhead is not performed
\newline - The dataset is imbalanced
\newline - They cannot differentiate in their model between emergency and attacks
\newline - The Augmentation in the number of attacks decreases the detection rate
\newline - High FNR and FPR
&
 - Dataset obtained from MIT-PHYSIOBANK \cite{LugovayaT.S2005}
\\
\midrule
(Hei et al., 2014) \cite{hei2014patient} 
& 
Supervised learning using the regression method by applying SVM 
& 
Detection of an abnormal dosage of insulin in insulin pumps 
&
 - Real time
\newline - High success rate in detecting single and chronic overdose attacks
\newline - The data are real, since they were collected through patients with diabetes
&
- Need software modification of insulin pumps
\newline- Overhead is not calculated well. Possibly saturate the memory of the insulin pumps since they need to collect three months of logs
\newline- This solution ensures the safety of a part of the insulin pump system and not the whole system
\newline- The model must be adapted for each patient, and a collection of 6 months of log files is necessary to create it, which means the patient is exposed to attacks during this period.
\newline- The value of the insulin dose that indicates an emergency is fixed for all patients
&
- Use of log files generated by the pump system from 4 patients
\newline- Not available
\\
\midrule
(Ahmad et al., 2018) \cite{ahmad2018securing}
&
LSTM and gesture recognition
& 
Prevent lethal insulin administration on patients
&
- Intuitive solution
&
- The patient must make a gesture to indicate if the amount of insulin in their body is good based on a symptom. However, what if the patient is young, and 
&
- Not available
\\
\bottomrule

\multicolumn{6}{r}{\footnotesize\itshape Continue on the next page}

\end{tabular*}
\end{table}

\begin{table}[width=\linewidth,cols=6,pos=H]
\caption*{Details of published studies that use ML for security purposes in Data Collection Level (continued)} \label{tbl3_1}

\begin{tabular*}{\linewidth}{ m{1.5cm} | m{1.5cm} | m{2cm} | m{3.2cm} | m{3.7cm} | m{1.5cm} }
\toprule
 &  &  &  & 
what if the patient becomes very sick and cannot perform gestures
\newline- The gesture sensor is also equipped with a wireless transmission module, which makes it vulnerable and must be secured
\newline- Their solution also requires a modification of the protocol adopted by the insulin pump
\newline- Their solution starts to work after three months of log file collection, which makes the system vulnerable during this period
\newline- The authors do not present details of their model, the results obtained, deployment of the IDS, and the analysis of their solutions
\newline- The solution proposed allows the protection of a single type medical equipment
 & \\
\midrule
(Rathore et al., 2017) \cite{rathore2017dlrt} 
& 
MLP, a DL approach
& 
Detection of fake glucose measurements and/or command on wireless insulin pump
&
- High accuracy
\newline- High reliability
\newline- Real time
\newline- Implemented on the chip, so it can be deployed on any device of the system using insulin pumps. These are the reason why the solution is reliable
\newline- Better recall compared with linear-SVM
&
- The authors propose to implement their solution on a chip that can be integrated in the IMD, which requires a modification of the device at the hardware and software level
\newline- They do not specify what is the action to take when an erroneous glucose measurement is detected
\newline- High space and time complexity compared with SVM
\newline- Non comparison is made with the non-linear SVM
\newline- The solution proposed allows the protection of a single type medical equipment
\newline- The emergency is not considered
&
- Dataset obtained from the public repository “UCI machine learning repository” \cite{VincentSigillito1990}
\\
\midrule
(Shobana, 2022) \cite{shobana2022towards}
& 
Unsupervised learning using the deep autoencoder
& 
Attacks detection using logs generated within Insulin Pump System 
&
- They use unsupervised learning, which avoids manual labeling of data
\newline- High scores (Accuracy, Precision, Recall and F1-measure)
&
- The attacks are simulated
\newline- Overhead and execution time are not calculated
\newline- The deployment of the model not mentioned

&
- Diabetes Data Set \cite{UMLRDDS2022}
\\
\bottomrule

\multicolumn{6}{r}{\footnotesize\itshape Continue on the next page}

\end{tabular*}
\end{table}

\begin{table}[width=\linewidth,cols=6,pos=H]
\caption*{Details of published studies that use ML for security purposes in Data Collection Level (continued)}\label{tbl3_2}
\begin{tabular*}{\linewidth}{ m{1.5cm} | m{1.5cm} | m{2cm} | m{3.2cm} | m{3.7cm} | m{1.5cm} }
\toprule

&

&

&

&
- The emergency is not considered
\newline- The solution proposed allows the protection of a single type medical equipment
&

\\
\midrule

(Rathore et al., 2019) \cite{rathore2019novel} 
& 
Supervised learning using LSTM
& 
Prevention of stimulation strategies attacks on DBS
&
- Real time
\newline- The loss value is low
&
- The authors did not specify where the solution is deployed
\newline- Accuracy not mentioned
\newline- Overhead not calculated
\newline- The attacks are simulated and are not real
\newline- The emergency is not considered
\newline- The solution proposed allows the protection of a single type medical equipment

&
- Data obtained from Physionet \cite{Goldberger2000}

\\
\midrule

 (Hei et al., 2010) \cite{hei2010defending}
 &  
 Linear and non-linear SVM
 & 
 Prevent resource dilapidation on IMD
 &  
 - High accuracy
\newline-  Real time
\newline- Add an extra layer of security to IMD
\newline-Solution designed for multiple medical devices
 & 
 - The solution is implemented on an external device (cell-phone) that can be stolen, lost or simply forgotten by the patient
\newline- Need software modification of the medical device
 \newline- At each authentication request, the IMD forwards this request to the cell-phone, which causes the IMD to consume energy and must be taken into consideration
\newline- This scheme is not designed to support emergencies
\newline- The protocol used between the cell-phone and the IMD is not secure
\newline- Their solution is not holistic since it is tested against only one type of vulnerable device
\newline- The patient must make a decision if the classification is not accurate
 & 
 - Not mentioned
\newline- Not available
 \\
\midrule
(Newaz et al., 2019) \cite{newaz2019healthguard} 
& 
ANN, DT, RF and KNN
& 
Detection of malicious activity in the smart healthcare system
&
- High accuracy and f1-score
\newline- Computation and detection time considered

&
- Overhead is not calculated
\newline- The FPR metric is not mentioned
\newline- They did not specify where this solution is deployed
&
- Use different datasets 
\\
\bottomrule
\multicolumn{6}{r}{\footnotesize\itshape Continue on the next page}
\end{tabular*}
\end{table}

\begin{table}[width=\linewidth,cols=6,pos=H]
\caption*{Details of published studies that use ML for security purposes in Data Collection Level (continued)}\label{tbl3_3}
\begin{tabular*}{\linewidth}{ m{1.5cm} | m{1.5cm} | m{2cm} | m{3.2cm} | m{3.7cm} | m{1.5cm} }
\toprule
&
&
&
- The systems proposed can determine if the alert generated is due to a disease or not
\newline- The dataset contains data of a healthy person as well as a diseased patient
\newline Solution designed for multiple medical devices
&
&
obtained from various repositories
\\
\midrule
(Salem et al., 2021) \cite{salem2020markov} 
& 
RMSE, Markov chain
&
Anomaly detection from the data collected by biosensors in WBAN composed of sensors and LPU
&
- High TPR. 
\newline- Low FAR
\newline- Solution designed for multiple medical devices
\newline- The solution is deployed on LPU, which reduces the transmission cost
\newline- Emergency considered 
&
- The attack is simulated
\newline- The model cannot differentiate between disease and attack if abnormalities are 
detected
\newline- Overhead is not calculated
&
Public dataset containing real data obtained from Physionet website \cite{MoodyGB1996}
\\

\bottomrule
\end{tabular*}
\end{table}

\subsection{ML for Transmission Level Security}
The transmission of medical data between devices composing the IoMT and the server enables the remote healthcare system to continuously monitor and treat patients in real time. However, the sensitive nature of the data exchanged represents a high interest for cyber-attackers, who, in case of a successful attack, can cause severe repercussions for the patient, ranging from violation of privacy to death. In addition, the heterogeneous nature of the devices used increases the surface of attack, which requires the design of a secure architecture for the IoMT.

One such proposal was made by (Gao and Thamilarasu 2017) in their paper \cite{gao2017machine}. Their proposed solution involves using ML methods to detect attacks targeting connected medical devices by learning the normal behavior of the device and identifying any deviations from it. The ML model is deployed on an external device that monitors the network and alerts the patient if an anomaly is detected. To test the effectiveness of their solution, the authors used three datasets of different sizes generated by a Castilia simulator \cite{OMNeT_based_simulator} and evaluated the performance of DT compared to SVM and k-means. After conducting various tests, the authors found that DTs provided higher accuracy, generated fewer false positives, and had a faster training and prediction time.

In another work performed by (Al-Shaher and al., 2017) in \cite{al2017protect}, they proposed to protect the private healthcare system from known viruses, worms, spyware, and denial-of-service attacks by designing and implementing an Intelligent Healthcare Security System (IHSS). The IHSS integrates the firewall, network intrusion detection subsystem, and web filter. The IHSS is intended to enhance the capabilities of these network protection systems using artificial intelligence approaches. The authors use MLP activated by wavelet transform to classify network traffic. The intrusion detection subsystem uses a wavelet neural network to determine which type of attacks are occurring by solving the multi-class problem. In web filters, they use Wavelet Neural Network to detect malware. After evaluating their method, they obtained 93\% accuracy with two hidden layers and 90\% with one hidden layer. 

Other research group (He et al., 2019) proposed an IDS based on a stacked Autoencoder for anomaly detection in the Connected Healthcare System, as outlined in their publication \cite{he2019intrusion}. The method involves several stages of data processing, including mapping, discretization, and normalization, before feeding the data to the stacked Autoencoder. The Autoencoder is used to extract the relevant features, which the ML models then use to perform detection and classify the data as either an attack or not.
To evaluate the performance of their IDS solution, the authors collected a real dataset from patients and simulated various types of attacks, such as DoS, counterfeit attacks, temper attacks, and replay attacks. They compared the performance of different ML models, including SVM, NB, KNN, and XGBoost, using metrics such as accuracy, FPR, and FNR. After conducting several tests, the authors found that the XGBoost model achieved the best performance with 97.83\% accuracy, 2.35\% FPR, and 1.65\% FNR.

Other researchers (Newaz et al., 2020) in \cite{newaz2020heka} presented HEKA an IDS based on ML for personal medical devices (PMD). The traffic generated between the PMD and the smartphone is analyzed with a sniffer to detect possible attacks using ML. The n-gram extracts feature sent to the IDS composed of four ML (KNN, DT, RF, and SVM). The HEKA is tested against four types of attacks, including a Man-In-The-Middle (MITM), false data injection, Replay, and DoS individually, then combining MITM and false data injection, and finally, MITM and Replay. They use eight devices composed of four PMDs (iHealth Air Wireless Pulse Oximeter, blood pressure monitor, QuardioArm blood pressure monitor, and wireless weight scale). The final dataset is composed of 731 benign instances and 308 malicious instances. After the realization of the different tests, they obtained a score of 98.4\% accuracy and 98\% F1-score.

In a recent study conducted by (RM et al., 2020) \cite{rm2020effective}, a new approach was proposed to develop an IDS based on DL  to predict and classify cyber attacks in the IoMT using a unique IP address. The proposed methodology aims to reduce the number of features and instances required for the classification process. This was achieved by transforming the categorical data into numerical data using one-shot coding, then normalizing it to a value between 1 and 0. Then, the normalized data is reduced using principal component analysis (PCA) at the first level and gray wolf optimization (GWO) at the second level, extracting only the most important features. The reduced dataset is then ranked using a DL algorithm using deep neural networks (DNN).
To evaluate the effectiveness of their proposed solution, the authors used a dataset obtained from Kaggle containing data collected by a wireless sensor network. They compared the performance of their methodology with other commonly used ML algorithms, including KNN, NB, RF, and SVM, using the measures of accuracy, specificity, and sensitivity. After performing various tests, the study found that the proposed methodology improves IDS accuracy by 15\% and reduces learning time by 32\%. These improvements would enable timely alerts to be generated in the event of healthcare system intrusion detection.

In another work carried out by (Lee et al., 2021) in \cite{lee2021m}, they proposed an IDS using ML and multi-class classification for the healthcare IoT within the smart city. The authors used CNN as an ML method to classify the network events generated by different medical devices into four classes, namely (critical, informal, major, and minor). Before the data are fed to the model, the data are preprocessed by transforming the categorical data into numerical data, then normalizing the data to take values within the same range. To evaluate their model, the authors generated a dataset by collecting data from six medical devices and then used them to compare their model results with other ML models regarding AUC, F1-score, Precision, and Recall. After performing the different tests, the authors found that their CNN model produces better results than other ML methods.

In another work performed by (Salemi et al. 2021), they presented a novel approach for predicting Distributed Denial of Service (DDoS) attacks in healthcare systems, as opposed to merely detecting them \cite{salemi2021leaesn}. The authors demonstrated that DDoS attacks cause traffic data to become chaotic, which can be analyzed using the Lyapunov Expansions Analysis and the Echo State Network. To implement their approach, the authors first represented network traffic as time-series data and applied a simple exponential smoothing method to predict future traffic. They then calculated the time-series prediction error by subtracting the predicted data from the actual data, which served as the basis for DDoS attack analysis. The authors then utilized a recurrent neural echo state network to predict the time series and used the LEA-MA method to detect the DDoS attack. To evaluate their method's effectiveness, the authors tested it on the DARPA 98 dataset \cite{MITLincolnLaboratory1998} and used metrics such as precision, recall, and F1-score. Their experiments showed that their proposed method could efficiently predict DDoS attacks.

In the following, a distributed IDS within the IoMT systems are presented as a solution. Among these solutions, there is work made by (Thamilarasu and Odesile 2017) in \cite{thamilarasu2020intrusion}. They proposed an approach that utilizes mobile agents to conduct penetration testing and secure the medical equipment network. The proposed system is characterized by its hierarchical, autonomous, and distributed nature. Intrusion detection is performed using a regression algorithm at the medical equipment level and ML techniques at the network level. Mobile agents traverse from one node to another or within a cluster, collecting network activities or device data based on their role as network or device intrusion detection agents. At the end of an intrusion test, the mobile agent classifies the collected samples as voluntary, malicious, or suspicious. The mobile agent migrates to another node if the samples are classified as voluntary. However, if the samples are malicious, an alarm is generated, and the data is sent to the cluster head. If the samples are suspicious, a request for intervention is sent to the cluster head. The cluster head deploys a special agent to collect data from the network or medical equipment of the entire cluster. The collected data are then tested to determine whether they are benign or malicious. Additionally, the system incorporates a security mechanism to detect intrusion at the cluster head level, which is achieved by using the cluster head agent that performs anomaly detection while traversing the cluster head network. The authors evaluated their proposed solution using OMNeT Castalia 3.2 simulator \cite{OMNeT_based_simulator} and tested five ML algorithms (SVM, DT, NB, KNN, and RF) to detect anomalies at the network level. Based on the accuracy, Cost Ratio, Feedback Reliability, training time, Total Rank Score, and Energy Overhead, DT produced the best results. Moreover, to detect anomalies in devices, the authors employed a cubic model of polynomial regression that balanced accuracy, overfitting, and computational resources. Finally, the proposed system was tested in a simulated hospital network topology and showed high accuracy, low overhead, and scalability.

Furthermore, in another study reported by (Begli et al. in 2019), a framework for securing remote healthcare systems was proposed \cite{begli2019layered}. Given the distributed nature of such systems, the authors employed multiple agents, each categorized based on their energy consumption and sensitivity to security risks. The resulting framework consists of three agent classes: sensors, smartphones, and databases. The first class, comprising sensors, utilizes a non-linear SVM-based anomaly detection approach, which consumes less energy and is effective for the limited data available from sensors. The second class, including smartphones with higher energy autonomy, uses a misuse-based intrusion detection system. The third and most critical class, which involves databases storing sensitive patient information, uses a hybrid detection system based on anomaly and misuse detection methods, as it is more prone to attacks and therefore has the highest security requirements. The authors evaluated the framework using the NSL-KDD dataset \cite{KDD22CONF} to detect various attacks, including DoS and user-to-root. The results demonstrated the effectiveness of the proposed framework, with efficient execution time, low energy consumption, high accuracy, and a low number of false positives.

In more recent studies, researchers explore other architectures that can be used in the context of IDS in IoMT systems: fog-cloud, SDN, and federated learning. In this regard, the work carried out by (Alrashdi et al., 2019) in \cite{alrashdi2019fbad} developed a framework for detecting attacks in fog nodes \cite{alrashdi2019fbad}. The authors used Online Sequential Extreme Learning Machine (OS-ELM) due to its fast learning speed. However, since OS-ELM produced inconsistent results, the authors employed a set of OS-ELM and employed majority voting to determine the presence of an anomaly. Prior to applying the ensemble of Online Sequential Extreme Learning Machine (EOS-ELM), the authors preprocess the data by converting discrete values into numerical values. Subsequently, they utilized the Information Gain algorithm and voting method to select features, then normalized the chosen attributes to values between 0 and 1. To evaluate and test their framework, the authors utilized NSL-KDD \cite{KDD22CONF}. After numerous tests, the authors found that EOS-ELM outperforms extreme learning machine, OS-ELM, and ML regarding accuracy, detection rate, and FPR. 

A recent study made by Kumar et al. (2021) in \cite{kumar2021ensemble} proposed an IDS that utilizes ensemble learning and a fog-cloud architecture to detect cyber-attacks in IoMT networks. The system preprocesses traffic data by converting categorical values into numerical values, replacing missing values with the mean of the corresponding feature values, and selecting the relevant features for intrusion detection using the correlation coefficient method. The numerical values are then normalized using the min-max technique to ensure they fall within a specific range.
The system uses a learning set consisting of NB, DT, and RF algorithms, which produce three prediction outputs. These outputs are then fed to XGBoost to produce the final output using majority voting. When an intrusion is detected, the administrator is alerted. The framework is deployed using a fog-cloud architecture, which utilizes Software as a Service at the fog level and Infrastructure as a Service at the cloud level.
The authors evaluated their framework using the Ton-IoT dataset \cite{NourMoustafa2019}, representing data collected from heterogeneous and large-scale IoT networks. The evaluation metrics were accuracy, detection rate, precision, FAR, and F1-score. The results show a detection rate of 99.98\%, an accuracy of 96.35\%, and a reduction of up to 5.59\% of the FAR. These results surpassed those of previous studies that used IDS.

In another research made by (Gupta et al., 2022) in \cite{gupta2022cybersecurity}, they proposed to use of deep hierarchical stacked neural networks to detect attacks that would attempt to modify the data flow, including meta-information that transits between the gateways and the edge cloud and between the edge cloud and the core cloud within multi-cloud healthcare systems called MUSE. This method includes reusing the edge cloud's trained layers to merge them and form a pre-trained model at the core cloud level. The tests were performed on three different datasets: UNSW-BOT-IoT and UNSW15 \cite{NourMoustafa2019} and one generated by the authors. A comparison was made with the method that does not reuse the trained layers of the edge cloud. The results show that the solution proposed by the authors improves the training efficiency and accuracy with a rate that varies between 95-100\% and reduces the training time by 26.2\%.

\begin{table}[width=\linewidth,cols=6,pos=H]
\caption{Details of published studies that use ML for security purposes in Transmission Level}\label{tbl4}
\begin{tabular*}{\linewidth}{ m{1.5cm} | m{1.5cm} | m{2cm} | m{3.2cm} | m{3.7cm} | m{1.5cm} }
\toprule
Ref  & Methods & Application & Advantage & Limitations & Dataset
\\
\midrule
(Gao and Thamilarasu, 2017) \cite{gao2017machine}  
& 
Different ML (DT, SVM, K-MEANS)
& 
Intrusion detection system for connected medical devices
& 
- High accuracy for DT
\newline- Low number of false positives
\newline- Their solution does not require any modification of software or hardware
\newline- Real time
& 
- The solution is implemented in external equipment that can be stolen, lost or forgotten
\newline- The external equipment where the solution is implemented must also be secured
\newline- In case of attack detection, there is no measure to stop the attack
\newline- It does not allow the detection of attacks emanating from an internal environment
\newline Lightweight solution not considered
\newline Synthetic datasets
\newline- Black-box model
& 
- The dataset is generated using the Castalia simulator \cite{OMNeT_based_simulator}
\newline- Not available
\\
\midrule

(Al-Shaher and al., 2017) \cite{al2017protect}
& 
MLP activated by wavelet transform and wavelet neural network
& 
Proposal to protect the private health system from cyberattacks by designing and implementing an IHSS
& 
- High accuracy
\newline- Real time detection
\newline- The proposed detection engine can be integrated into web filter, intrusion detection, and firewall
&
- Not all metrics are used
\newline- It does not allow the detection of attacks emanat
ing from an internal environment 
\newline- Overhead not considered
\newline- Black-box model
\newline- No measure to protect privacy 
&
- Not mentioned
\\
\midrule

(He et al., 2019) \cite{he2019intrusion}
&
Stacked Autoencoder to extract features and XGBoost to perform classification
&
Proposal of an IDS based on a stacked Autoencoder for anomaly detection in the connected healthcare system
&
- High accuracy
\newline- Low FPR and FNR
\newline- Real time
\newline- Lightweight solution 
&
- The deployment of the solution is not mentioned
\newline- It does not allow the detection of attacks emanating from an internal environment
\newline- Overhead is not calculated
\newline- The dataset is imbalanced
\newline- Black-box model
\newline- No measure to protect privacy 
&
- Self-generated datasets were collected from patients, and attacks were simulated
\\
\midrule

(Newaz et al., 2020) \cite{newaz2020heka}
&
n-grame for features extraction and four ML (KNN, SVM, RF, DT)
&
Detection of cyber-attack on PMDs
&
- High accuracy
\newline- High F1-score
\newline- System is tested against the different types of attacks using different types of PMD
\newline- Scalable.
\newline- Perform passive analysis, which is effective in terms of performance overhead
\newline- Real time
&
- Not all performance metrics are used
\newline- Detection time is not considered
\newline- The combination of ML methods deteriorates the performance of HEKA

&
- The data are generated from different real devices
\newline- The dataset is not publically available

\\
\bottomrule
\multicolumn{6}{r}{\footnotesize\itshape Continue on the next page}
\end{tabular*}
\end{table}

\begin{table}[width=\linewidth,cols=6,pos=H]
\caption*{Details of published studies that use ML for security purposes in Transmission Level (continued)}\label{tbl4_1}
\begin{tabular*}{\linewidth}{ m{1.5cm} | m{1.5cm} | m{2cm} | m{3.2cm} | m{3.7cm} | m{1.5cm} }
\toprule
&
&
&
&
- Their solution is only tested against the 'just work' authentication mechanism of the Bluetooth and is not tested against the 2 others which are passkey entry and out of band methods
\newline- It does not allow the detection of attacks emanating from an internal environment
\newline- The data are collected from healthy persons
\newline- Black-box model
&
\\
\midrule
(RM et al., 2020) \cite{rm2020effective}  
& 
PCA and GWO to reduce the dimensionality of data and DNN to classify data
& 
Intrusion detection system in IoMT system
& 
- Increase in accuracy by 15\% and decrease in time complexity by 32\% comparing with traditional ML approaches
\newline- The reduction in data dimensions, reduce the time complexity and increase the accuracy
\newline- The grey wolf optimization allows reducing the drawback of the local minimum
& 
- The dataset used is not designed for IoMT
\newline- The implementation of this solution is not mentioned
\newline- Overhead is not calculated
\newline- Black-box model
\newline- No measure to protect privacy 
& 
- Kaggle intrusion data samples were collected from Kaggle open-source center

\\
\midrule

(Lee et al., 2021) \cite{lee2021m}
&
CNN
&
IDS designed for healthcare IoT in the smart city
&
- High accuracy
\newline- Real time
\newline- Low computational overhead
\newline- Multi-class classification
\newline- Different medical devices used to collect data
& 
-The class major has a low accuracy rate
\newline- Black-box model
\newline- No measure to protect privacy 
 & 
- Dataset generated from six medical devices
\newline- Not available

\\
\midrule

(Salemi et al., 2021) \cite{salemi2021leaesn}
&
Lyapunov Exponent Analysis and Echo State Network
&
Prediction of DDoS attacks in multimedia-based healthcare systems
&
- High accuracy
\newline- Real time
\newline- Can predict the DDoS attack instead of detecting it
&
- Overhead is not calculated
\newline- Old dataset, and it is not specific for IoMT
\newline- Black-box model
\newline- No measure to protect privacy 
&
DERPA 98 dataset, publically available \cite{MITLincolnLaboratory1998}

\\
\midrule

(Thamilarasu and Odesile, 2017) \cite{thamilarasu2020intrusion}  
& 
Mobile agent-based intrusion detection system using ML and regression algorithms
& 
Securing the network of connected medical devices
& 
- High detection accuracy
\newline- Low overhead
\newline- Detection intrusions are performed at the network and device level 
\newline- Systems are scalable, hierarchical, distributed, fault-tolerant, and autonomous
& 
- Consume energy at the sensor agent level
\newline- The patient data are sent in some cases to the cluster head, which represents a possible violation of privacy
\newline- Black-box model
& 
- Data are generated using OMNeT Castalia simulator \cite{OMNeT_based_simulator}
\newline- Not available

\\
\bottomrule
\multicolumn{6}{r}{\footnotesize\itshape Continue on the next page}
\end{tabular*}
\end{table}

\begin{table}[width=\linewidth,cols=6,pos=H]
\caption*{Details of published studies that use ML for security purposes in Transmission Level (continued)}\label{tbl4_3}
\begin{tabular*}{\linewidth}{ m{1.5cm} | m{1.5cm} | m{2cm} | m{3.2cm} | m{3.7cm} | m{1.5cm} }
\toprule

(Begli and al., 2019) \cite{begli2019layered}
&
IDS based on supervised learning using non-linear SVM and misuse detection systems
&
Detection of common attacks including DoS and user to root in remote healthcare systems
&
- High detection rate for hybrid and anomaly detection systems
\newline- Real time
&
- Low detection rate for misuse detection systems
\newline- The dataset used is not specifically made for the IoMT
\newline- Overhead not calculated
\newline- Not all metrics are used
\newline- Rule number 2: This rule does not allow the detection of new vulnerabilities and zero-day attacks
\newline- Black-box model
\newline- No measure to protect privacy 
&
NSL-KDD dataset, publically available \cite{KDD22CONF}

\\
\midrule
(Alrashdi et al., 2019) \cite{alrashdi2019fbad}  
& 
EOS-ELM
& 
Framework for attack detection in the Fog node
& 
- High detection rate
\newline- Low latency
\newline- Real time
\newline- The system is distributed
& 
- Overhead is not calculated
\newline- No measure to protect privacy
\newline- The dataset used is not specific for IoMT
\newline- Not all metrics are used
\newline- Black-box model
& 
- NSL-KDD dataset \cite{KDD22CONF}

\\
\midrule

(Kumar et al., 2021) \cite{kumar2021ensemble} 
& 
NB, DT, RF and XGBoot
&
Cyber-attack detection in IoMT networks using fog-cloud architecture
&
- Distributed solution
\newline- High detection rate
\newline- High accuracy
\newline- Low FAR
&
- Training time for ensemble learning is higher than the different ML used alone
\newline- The dataset used is not specific for IoMT 
\newline- NB present the worst result for different metrics used to evaluate the ML model
\newline- Black-box model
\newline- No measure to protect privacy 
&
Ton-IoT dataset, publicly available \cite{NourMoustafa2019}
\\
\midrule

(Gupta et al.,2022) \cite{gupta2022cybersecurity}
&
Deep hierarchical stacked neural networks
&
IDS to detect modification in data flow within multi-cloud healthcare systems
&
- High accuracy
\newline- Low execution time
& 
- The model needs to be trained with data at the core cloud, which could impact data privacy 
\newline- Overhead is not calculated
\newline- Black-box model
 & 
- UNSW-BOT-IoT and UNSW15 \cite{NourMoustafa2019} datasets
\newline- Generated dataset

\\
\midrule

(Hameed et al.,2022) \cite{hameed2022whte}
&
Weighted Hoeffding Tree Ensemble
&
Fog-based IDS for the industrial IoMT
&
- Lightweight solution 
\newline- Their solution can be deployed on edge and fog  
\newline- High accuracy 
& 
- The dataset used is not designed for IoMT
\newline- Black-box model
\newline- No measure to protect privacy
 & 
- NSL-KDD \cite{KDD22CONF} and ToN-IoT \cite{NourMoustafa2019} datasets

\\
\bottomrule
\multicolumn{6}{r}{\footnotesize\itshape Continue on the next page}
\end{tabular*}
\end{table}

\begin{table}[width=\linewidth,cols=6,pos=H]
\caption*{Details of published studies that use ML for security purposes in Transmission Level (continued)}\label{tbl4_4}
\begin{tabular*}{\linewidth}{ m{1.5cm} | m{1.5cm} | m{2cm} | m{3.2cm} | m{3.7cm} | m{1.5cm} }
\toprule

(Khan and Akhunzada, 2021) \cite{khan2021hybrid} 
& 
CNN to extract feature and LSTM to classify data
& 
Hybrid deep learning-based model for malware detection in the IoMT deployed at the SDN plane application level
& 
- High detection accuracy
\newline- Speed efficiency
& 
- This framework is prone to a single point of failure
\newline- Black-box model
\newline- No measure to protect privacy
& 
Use of publicly available datasets. However, they do not mention it

\\
\midrule

(Wahab et al., 2022) \cite{wahab2022ai}
& 
hybrid model combining LSTM and GRU
& 
IDS based on ML to secure the IoMT architecture based on SDN against cyber threats
& 
- High accuracy, precision and F1-score
\newline- Fast execution
& 
- The dataset used is not specific to IoMT systems
\newline- The solution is centralised, which suffers from a single point of failure
\newline- Overhead not calculated 
\newline- No measure to protect privacy
\newline- Black-box model
& 
CICDDoS2019 dataset \cite{sharafaldin2019developing}

\\
\midrule
(Schneble and Thamilarasu, 2019) \cite{schneble2019attack} 
& 
Federated Learning
& 
Securing the medical cyber-physical systems
& 
- The system is distributed and scalable
\newline- High detection accuracy
\newline- Low FPR
\newline- Real time
\newline- Low communication overhead
\newline- Protects patient privacy by transmitting only the model instead of the patient data
& 
- The model of ML is not protected during transmission
\newline- The minimum and the maximum number of mobile devices that should be taken into consideration is not determined
\newline- They did not specify how the medical staff is notified in case of anomaly detection
\newline- Black-box model
& 
- MIMIC Dataset obtained from Physionet,  publically available \cite{Johnson2016}

\\
\midrule

(Singh et al., 2022) \cite{singh2022dew}
& 
HFL based on hierarchical long-term memory
& 
IDS for distributed dew servers of the IoMT system
& 
- High accuracy, precision, f-score and recall
\newline- Reduced computation cost
\newline- Data privacy preserved
& 
- Dataset not made for IoMT applications
\newline- No measures to protect models' transmission
\newline- Black-box model
\newline- Execution time not considered
& 
- TON-IoT \cite{NourMoustafa2019} and NSL-KDD \cite{KDD22CONF} datasets

\\
\midrule

(Khan et al., 2022) \cite{khan2022xsru}
& 
Simple Recurrent Units with skip connections and Local Interpretable Model-Agnostic Explanations
& 
IDS for IoMT network deployed at gateway and router levels
& 
- High accuracy, Precision, Recall, and F-measure
\newline- Low computation cost
\newline- Time efficiency 
\newline- Model explained and interpreted 
& 
- Dataset used is not specific for IoMT systems
\newline- The solution is centralised, which is prone to a single point of failures 
\newline- No measure to protect privacy
& 
ToN-IoT \cite{NourMoustafa2019} dataset

\\
\bottomrule
\multicolumn{6}{r}{\footnotesize\itshape Continue on the next page}
\end{tabular*}
\end{table}

\begin{table}[width=\linewidth,cols=6,pos=H]
\caption*{Details of published studies that use ML for security purposes in Transmission Level (continued)}\label{tbl4_5}
\begin{tabular*}{\linewidth}{ m{1.5cm} | m{1.5cm} | m{2cm} | m{3.2cm} | m{3.7cm} | m{1.5cm} }
\toprule

(Hady et al., 2020) \cite{hady2020intrusion}  
& 
They compare the results of different ML methods
& 
IDS for healthcare using medical and network data
& 
- High accuracy and AUC
\newline- Real time
\newline- Datastes designed for IoMT application 
& 
- The dataset contains only two types of attacks
\newline- The dataset is imbalanced
\newline- Computation overhead is not considered 
\newline- The IDS are located between the gateway and server. However, the attack can occur between medical equipment and getaway
\newline- Black-box model
& 
Dataset publically available \cite{DTST22CONF}

\\
\bottomrule
\end{tabular*}
\end{table}

The proposed IDS solution at the fog level has the advantage of being close to the IoT devices and therefore offers a rapid response, decentralization and preserves data privacy. However, the fog level faces an increase in the amount of data arriving at the fog level, which requires a lightweight solution. For this purpose, (Hameed et al.,2022) in \cite{hameed2022whte} proposed to use an incremental ensemble learning method called Weighted Hoeffding Tree Ensemble system consisting of an incremental learning classifier for the industrial IoMT. Tests on NSL-KDD \cite{KDD22CONF} and ToN-IoT \cite{NourMoustafa2019} datasets and comparison with single incremental classifiers and Bagging Hoeffding Tree ensemble algorithms show that the proposed solution is lightweight and presents a trad-off between accuracy and overhead (CPU, memory and time) and outperforms the results of previous studies.

Other works have proposed SDN as an architecture for ML-based IDS within IoMT systems. Among them is the work led by (Khan and Akhunzada, 2021) in \cite{khan2021hybrid}, which proposed a hybrid model based on DL for malware detection in IoMT deployed at the SDN plane application level.
The system proposed by the authors consists of feature extraction using CNN, and then LSTM is used to classify the data as malware. The authors used the current state-of-the-art IoT malware publicly available dataset to evaluate their model. In addition, they compared their model with the constructed hybrid DL-driven. After performing various tests, the authors found that the proposed model outperformed the other methods regarding detection accuracy and speed efficiency.

In another study, (Wahab et al., 2022) in \cite{wahab2022ai} proposed to use IDS based on ML to secure the IoMT architecture based on SDN against cyber threats. For this purpose, the authors propose using a hybrid model combining the LSTM and GRU deployed at the SDN control plane. The tests were performed on the CICDDoS2019 dataset \cite{sharafaldin2019developing}, which was subjected to a preprocessing consisting of eliminating the missing value, transforming the non-numeric values into numeric values, applying the one-hot encoding to output label, and then using MinMaxScaler for data normalization. They compared the results with other classifiers: cu-GRU, DNN, and cu-BLSTM, as well as other previous studies. The findings indicate that their solution outperforms other models and prior studies in terms of accuracy, precision, F1-score, and execution time.

The various IDS solutions proposed to improve the security of the IoMT network do not include measures to protect patient privacy. In this perspective, the research realized by (Schneble and Thamilarasu, 2019) in \cite{schneble2019attack} proposed to use an IDS based on ML and federated learning to secure medical cyber-physical systems composed of sensors, mobile devices, and servers. The distributed and scalable nature of this system makes it more effective in protecting patient data privacy.
The proposed IDS system follows a process where mobile devices first register with the server and are assigned to a cluster based on their health history. This cluster is associated with a federated model stored on the server. Each mobile device downloads the federated model, trains it, and updates it using patient data. The server then selects some or all of the mobile devices that compose the cluster and asks them to send their updated model. The server then calculates the average weights and biases of the received models to update its federated model, which mobile devices can download. This process continues until the model converges.
Mobile devices can be in two modes: testing and learning. In learning mode, the mobile device can predict and send its updated model to the server. In the testing mode, the mobile device can only predict the new data and does not send the model to the server, saving communication costs. The proposed system allows the detection of anomalies, such as a value of an attribute that exceeds its usual value range or has an unexpected correlation with other attributes. In these cases, an alert is generated on mobile devices, allowing the nursing staff to react.
The system was tested using the MIMIC dataset from Physionet \cite{Johnson2016}, with the addition of simulated attacks such as DoS, data modification, and injection. The performance metrics used to evaluate the system were detection accuracy, FPR, recall, F1-score, training time, and communication overhead. The tests produced high detection rates and low false positives, with training times equivalent to or better than using a single ML. Additionally, increasing the number of patients does not affect the training time, which improved the accuracy and decreased the FPR by obtaining more data.

In another study made by (Singh et al., 2022) in \cite{singh2022dew}, the author proposed a solution for intrusion detection at the IoMT networks level using Hierarchical Federated Learning (HFL) based on hierarchical long-term memory. Their solution is deployed on the distributed dew servers of the IoMT framework, with a backend supported by cloud computing at the edge layer. The proposed HFL solution aggregates models from different entities  that compose the healthcare institution at the dew-servers level to obtain local models. These local models are then aggregated at the cloud computing level to obtain the global model, which is redistributed to the different dew servers participating in the learning process. This iterative process is repeated until the global model achieves a high accuracy.
The authors evaluated their solution on TON-IoT \cite{NourMoustafa2019} and NSL-KDD \cite{KDD22CONF} datasets. They preprocess the datasets by removing unnecessary features, converting non-numeric values to numeric, and applying one-hot encoding to categorical values before normalizing the data to ensure they are represented in the same range. Finally, they applied PCA for dimensionality reduction. The results showed high accuracy, precision, recall, and f-score compared to Recurrent Neural Network (RNN), Gated Recurrent Unit (GRU), LSTM, and the previous study, with minimized computation costs.

The previous ML models lack interpretability and do not explain how they detect attacks, making them black-box models. However, (Khan et al., 2022) presented a solution to this issue in their study \cite{khan2022xsru}. They utilized Simple Recurrent Units with skip connections to avoid the vanishing gradient problem for detecting network attacks in the IoMT, which was deployed at the gateway and router levels. To increase trust in their model, the authors employed Explainable AI (XAI) design using the Local Interpretable Model-Agnostic Explanations model to interpret and explain how the model classifies the data. The evaluation was performed on the ToN-IoT \cite{NourMoustafa2019} dataset, which underwent preprocessing such as labeling categorical data, normalizing data to the same scale, and applying PCA for dimensionality reduction. The results indicate that the proposed solution by the authors is more efficient than two RNN variants: LSTM and GRU, in terms of Accuracy, Precision, Recall, and F-measure. Additionally, the proposed solution has a reduced computation cost compared to previous studies. Furthermore, XAI revealed that the basic category of IoT device features significantly impacts data classification.

The datasets used to evaluate the ML models in the above solutions consist of either network or medical data. The research performed by (Hady et al., 2020) in \cite{hady2020intrusion}, proposed an IDS for healthcare that uses medical and network data. For this purpose, the authors developed an architecture that allows the creation of a dataset containing medical and network data and simulated MITM attacks to perform two types of attacks: spoofing and data alteration. The generated dataset \cite{DTST22CONF} is used to test and evaluate different ML methods: SVM, KNN, RF, and ANN, with the following metrics accuracy, AUC, and time of execution for training and testing. After performing the different tests, the authors found that their system, which combines medical and network data increased the effectiveness of ML methods by 7 to 25\% for the detection of threats in health monitoring systems in real time.

A literature review reveals that current proposals for using IDS based on ML in the IoMT environment have mainly explored centralized architecture solutions \cite{gao2017machine, al2017protect, he2019intrusion, newaz2020heka, rm2020effective, lee2021m, salemi2021leaesn, khan2022xsru, hady2020intrusion, khan2021hybrid, wahab2022ai}. However, some researchers have also investigated distributed solutions \cite{thamilarasu2020intrusion, begli2019layered}. Incorporating IDS and blockchain technology to achieve a decentralized architecture in the IoMT setting presents a promising direction for future research. Since it can bring numerous benefits to the IoMT environment, for example, it can enhance security by removing the risk of a single point of failure, making it difficult for malicious actors to alter data. The immutability of blockchain technology also ensures that data is recorded permanently, providing a trustworthy record of transactions. Decentralization can also  improve traceability and reduce dependency on intermediaries, leading to more efficient and transparent processes. However, several challenges are also associated with using blockchain technology in the IoMT environment. Scalability remains a significant concern, as current processing speed and storage capacity limitations may not be suitable for large-scale applications. Additionally, the energy consumption associated with proof-of-work consensus algorithms can negatively impact the environment. Finally, the lack of standardization across different blockchain platforms can limit interoperability between them.

 The future solution must prioritize protecting patient privacy, provide interpretable and explainable ML models, and apply appropriate learning and testing processes using a custom-designed dataset in an IoMT environment. Adherence to these requirements will ensure the development of a secure, transparent, and trustworthy IDS solution for the IoMT setting, which is critical in today's increasingly connected healthcare ecosystem.

 \subsection{ML for Storage Level Security}
Medical data received from sensors are centralized in a medical server, which the medical staff can access for analysis. These stored data are of two types: the EMR and the EHR. EMR stores a patient's medical and treatment history in a single place and makes it accessible at a single hospital. While EHR focuses on the patient's general health, it can store and transmit the patient’s health data, such as patient history, medication, test results, and demographics \cite{PeterGarrettandJoshuaSeidman2011, CascioTeresa}. It is necessary to secure access to these data to preserve patients' privacy, protect the confidentiality of medical data, and guarantee their availability and integrity to make an accurate diagnosis. 

In this perspective, (Boxwala et al., 2011) in \cite{boxwala2011using} proposed to use statistics and ML to identify suspicious access in EHR access logs. The authors used Logistic Regression (LR) and SVM to classify new access as suspicious with ranking. The high-scoring event is investigated first by the privacy officers. To create the model based on LR and SVM,
the authors used the privacy agent to label the selected events as suspicious or appropriate using an iterative refinement process. Then they trained the model using 10-fold cross-validation. The authors used sensitivity, AUC and compared their model with the rule-based technique to evaluate their model. After performing several tests, the authors obtained more than 0.90 of AUC and more than 0.75 of sensitivity. They find that using a method based on statistics and ML to detect suspicious access in EHR is possible and is more effective than a rule-based technique. For the same purpose, a different approach is proposed by (Menon et al., 2014) in \cite{menon2014detecting} for detecting privacy violations resulting from inappropriate access to EHR. The authors use an approach inspired by collaborative filtering 
for inappropriate access detection, where the objective is to predict a label for a pair of entities interactions. Their solution incorporates explicit and latent features for staff and patients, allowing for the generation of a fingerprint customizer for users based on previous access history. To evaluate the model, the authors used two datasets named "hospital" and "amazon" \cite{Amazon_Access_Data_Competition} using the following metrics: RMSE, the area under curves, and precision-recall curves, then they compared the results obtained with three ML algorithms: linear regression, LR, and SVM. After performing the different tests, the authors improved the performance considerably over the other approaches and detected inappropriate access.

Furthermore, the work performed by (Malin and Bradley, 2014) in \cite{Malin2014} proposed an unsupervised learning model for insider threat detection in a collaborative environment using access logs called a community-based anomaly detection system. The approach proposed by the authors is hybrid; they use singular value decomposition, a special case of PCA, to infer communities from relational networks of users, and then they use KNN to create a set of nearest neighbors. The created model detects anomalous users by identifying users who have diverged from typical communication behaviors. To evaluate their model, the authors used two datasets: a six-month collection of access logs from an actual EMR and another dataset that reports the editorial board composition for a set of journals over five years. After running the different tests, the results showed that their model could detect the simulated user with high accuracy, outperforming other anomaly detection models.

Another work carried out by (Marwan and al., 2018) in \cite{marwan2018security}, presented a new approach to secure image data processing in the cloud environment based on ML. Their method consists of segmenting the image into four distinct parts depending on pixel intensity level using Fuzzy C-Means Clustering (FCM) and SVM. The FCM is utilized for extracting color features at the pixel level. These features are fed to the SVM to be classified into different regions, allowing storage of the image in the cloud in a segmented format. The authors have also proposed a 3-layer architecture instead of the traditional 2-layer architecture by introducing a CloudSec module that allows the encryption of data in transit using the HTTPS/Secure Socket Layer. The CloudSec module also allows for restricting access to the data and detecting the misuse of cloud resources by using an access control mechanism.

In a work performed by (Sicuranza and Paragliola, 2020) in \cite{sicuranza2020ensuring}, they proposed a hybrid IDS for cyber-attack detection against EHR. 
The proposed system uses agents deployed within the monitored IT infrastructure. They are responsible for collecting, normalizing, and performing security analysis on the logs collected from the local level. Then these agents generate events that are sent to the IDS for analysis. The IDS comprises a misuse detection module and an anomaly detection module. The misuse detection module is rule-based, effectively detecting the well-known attack signature. The anomaly detection module allows the detection of zero-day attacks. Anomaly detection uses three classifiers: DT, Neural Network, and k-means. The results of these classifiers are sent to the voting system to improve the accuracy of each classifier. In addition, an expert system module is designed to resolve any potential conflict between the presence/absence of attacks as determined by the abuse detection module and the anomaly detection voting system. 
A dataset was generated by monitoring the Italian EHR system to test the proposed model. Three separate attacks on the EHR systems were used to test the misuse and anomaly detection modules. The results demonstrate the efficiency of the proposed solution.

\begin{table}[width=\linewidth,cols=6,pos=H]
\caption{Details of published studies that use ML for security purposes in Storage Level}\label{tbl5}
\begin{tabular*}{\linewidth}{ m{1.5cm} | m{1.5cm} | m{2cm} | m{3.2cm} | m{3.7cm} | m{1.5cm} }
\toprule

Ref 
& 
Methods
& 
Application
& 
Advantage
& 
Limitations
& 
Dataset
\\
\midrule

(Boxwala et al., 2011) \cite{boxwala2011using} 
& 
LR and SVM
& 
Help institutions detect suspicious access to electronic health records
& 
- Good AUC
\newline- Good sensitivity
& 
-Their method requires the intervention of privacy agents, who are not always present, nor have the necessary knowledge to differentiate between what may or may not constitute a violation
\newline- The solution is not real time
\newline- The problem has only been addressed by one institution
\newline- The solution is not fully automated since labeling dataset is performed manually
\newline- Not all metrics are used
\newline- Not real time

&
Not available
\\
\midrule

(Menon et al., 2014) \cite{menon2014detecting} 
& 
Collaboration filtering using latent and explicit features
& 
Detection of privacy breaches resulting from inappropriate access to EHR
& 
- The proposed model offers a significant improvement over supervised learning methods
& 
-Their approach implies the intervention of privacy officers, who may not be present and do not have the expertise to distinguish between violations and non-violations
\newline- The model is not real time
\newline- The solution is applied to one institution
\newline- They use a small training dataset because privacy officers do the labelling of data manually
\newline- Not all metrics are used
\newline- The model is not fully automated
&
- Use two datasets named «hospital» and «amazon»
\newline- The dataset «hospital», not available
\newline- The dataset «amazon»  is available at \cite{Amazon_Access_Data_Competition}
\\
\midrule
(Malin and Bradley, 2014) \cite{Malin2014}
& 
Unsupervised learning model using singular value decomposition and KNN 
& 
Detection of insider threat in collaboration environment using access logs, named community-based anomaly detection system
& 
- High degree of certainty in differentiating anomalous users from real users
\newline- Outperform other states of the art approaches
& 
- Experts are needed to validate the result (not fully automated)
\newline- The number of k cluster in KNN change from one system to another
\newline- This approach cannot detect attackers that imitate legitimate group behavior or legitimate behavior of another user
&
- the First dataset is a collection of six months of access logs from real electronic health records 
\newline- The second
\\
\bottomrule

\multicolumn{6}{r}{\footnotesize\itshape Continue on the next page}

\end{tabular*}
\end{table}

\begin{table}[width=\linewidth,cols=6,pos=H]
\caption*{Details of published studies that use ML for security purposes in Storage Level (continued)}\label{tbl5_1}
\begin{tabular*}{\linewidth}{ m{1.5cm} | m{1.5cm} | m{2cm} | m{3.2cm} | m{3.7cm} | m{1.5cm} }
\toprule
& 
& 

& 
& 
- Not real time
&
dataset is the report of the editorial board membership for a set of journals discipline over 5 years
\newline- Not available
\\
\midrule

(Marwan and al., 2018) \cite{marwan2018security}
& 
combination of SVM and FCM
& 
Securing image data processing in a cloud environment based on machine learning
& 
- Good alternative to data encryption
& 
- Small dataset (only two images)
&
Use of two images for testing purposes
\\
\midrule

(Sicuranza and Paragliola, 2020) \cite{sicuranza2020ensuring}
& 
Rule, DT, neural network and k-means
& 
Hybrid IDS for the detection of cyber-attack against the EHR system
& 
High accuracy for anomaly detection module
& 
- The time of detection is not calculated
\newline- The result of the misuse detection module is not reported
\newline- The solution is applied to one institution
\newline - Due to the requirement for an expert to resolve any conflicts in the presence or absence of attacks, the solution is not fully automated

&
The Dataset was produced by monitoring the Italian EHR system and simulate three different attacks (not available)
\\
\midrule

(McGlade and Scott-Hayward, 2019) \cite{mcglade2019ml}
& 
SVM and EMA
& 
Framework for confidentiality and availability issue detection in EMR systems
& 
- High detection accuracy
\newline- High recall
& 
- Not all metrics are used.
\newline- The integrity of the system EMR is not checked throw the ML method
&
- Synthea was used to simulate the representation of patient data in the FHIR Database by generating patient information such as names and addresses \cite{synthetichealth}
\\

\bottomrule
\end{tabular*}
\end{table}

In addition, in a study reported by (McGlade and Scott-Hayward, 2019) in \cite{mcglade2019ml}, they proposed a framework for detecting privacy and availability issues in EMR systems. The framework is based on ML and uses the SVM to detect privacy-related incidents and the Exponential Moving Average (EMA) to detect anomalies in message flow that may cause a denial of service. To test the framework, the authors have used synthetic data generated by the Synthea tool \cite{synthetichealth}, a synthetic patient population simulator. They have tested three ML algorithms on the dataset, namely SVM, KNN, and multinomial NB, to detect anomaly-related to the confidentiality of the EMR system and EMA for the detection of anomaly-related to the availability of the EMR system. After performing different tests, the authors found that SVM exhibits the best performance regarding accuracy and recall than the two other methods. They also find that EMA can successfully detect message surges, leading to a denial of services.

EHRs and  EMRs are critical systems for storing sensitive patient information. Ensuring this information's confidentiality, integrity, and availability are of utmost importance. The literature review reveals that most studies have concentrated on detecting unauthorized access to EHRs \cite{boxwala2011using, menon2014detecting, Malin2014}. Only one study has focused on the confidentiality and availability of EMRs \cite{mcglade2019ml}. However, there is a need for further research in ML to ensure the integrity of both EHRs and EMRs. Exploration in this area holds great promise and could be a valuable direction for future investigation.

\section{CHALLENGES AND LIMITATIONS} \label{CHALLENGES AND LIMITATIONS}
The exploration of the different solutions proposed in the literature using IDS based on ML for IoMT led us to identify the limitations and challenges of this approach at the different layers that compose IoMT as follows:

\subsection{Data Collection Level Security}

The deployment of ML models in medical equipment presents a challenge due to its various limitations. Three methods of deployment have been proposed, which include deployment of the ML model on medical equipment \cite{kintzlinger2020cardiwall}, deployment on a third-party device \cite{hei2010defending}, or deployment on a chip followed by integration into the medical equipment \cite{rathore2017dlrt}.

The deployment of ML models on resource-constrained medical devices can have consequences such as shortened battery life, which may require surgery for battery replacement and pose a risk to the patient. Deployment on third-party devices involves communication between the devices and medical equipment, which requires a modification at the software level of the medical device. However, medical device manufacturers do not permit such changes, and third-party devices must also be protected from potential attacks. Deployment on a chip and integration into the medical device involves software and hardware modifications.
Lightweight ML models that satisfy the limitations of sensors may be a promising area of research in addressing the challenge of deploying ML models on resource-limited medical devices. Additionally, the ML model must be secured to prevent data manipulation during the learning or testing phase that can compromise the results \cite{hameed2021systematic, Newaz2020a}.

Detecting anomalies in medical data can arise from various factors, including poor communication quality due to interference or malfunctioning sensors, medical emergencies that result in severe illness, and security attacks carried out by malicious entities. It is crucial to differentiate between these various sources of anomalies, as prompt identification and treatment are necessary for emergencies. 

The limited availability of public datasets generated explicitly for security purposes and containing medical data presents a challenge for researchers. Some have resorted to using existing medical datasets, such as PHYSIONET \cite{Johnson2016}, and modified specific values with the help of healthcare professionals to simulate attacks, leading to questions about the validity of their findings when applied in real-world scenarios. Other studies have used real medical equipment volunteers wear to collect health data. However, these results may need to be more representative as the volunteers may not have any underlying health conditions. Using simulators, such as CASTILIA \cite{OMNeT_based_simulator}, to generate medical data and attacks can also result in unforeseen difficulties during practical implementation. The acquisition of a high-quality, diverse, and representative dataset containing medical data collected from individuals with and without illnesses, and generated explicitly for security purposes, remains a significant challenge.

Another limitation is that the various IDSs proposed in the literature focus on a restricted number of medical devices to generate a medical dataset. However, when patients utilize multiple medical devices, it becomes imperative to have integrated solutions to detect intrusions from these different medical devices. The challenge lies in designing an IDS solution that can accommodate all connected medical devices, as each has specific data collection and communication methods.

Some solutions rely on the patient's ability to determine the occurrence of an attack. However, this approach needs to consider potential limitations such as age, incapacity, or emergency circumstances where the patient may be unable to make such a decision. Establishing effective protocols for communication and decision-making in the event of anomalous behavior detection poses a challenge.

The study \cite{kintzlinger2020cardiwall} has demonstrated that rule-based solutions for anomaly detection in medical data may achieve better results than ML-based anomaly detection systems because some medical attributes contain a range of values that, if exceeded, can be easily detected by implementing clear and defined rules. On the other hand, ML-based anomaly detection can detect anomalies by analyzing correlations between multiple medical attributes. This highlights the need for a comprehensive evaluation of rule-based and ML-based anomaly detection methods to determine their strengths and limitations in detecting anomalies in medical data. Further research should consider this aspect to improve the accuracy and reliability of anomaly detection in medical data.

The generalizability of a ML model constructed from a specific patient's medical dataset is limited, as what may be considered abnormal medical values for one patient may be considered normal for another patient \cite{salem2020markov}. Ensuring the accuracy of the ML models when applied to different patient populations with varying medical conditions and characteristics remains a challenge.

\subsection{Transmission Level Security}

In developing a Network-based IDS for the IoMT, the unique characteristics of the IoMT system must be considered, such as its distributed, mobile, dynamic nature and heterogeneous communication constraints.

When designing an ML-based Network-based IDS for IoMT security, it is essential to consider the evaluation metrics for the ML model, especially when the dataset is imbalanced. The use of metrics that give higher importance to the minority classes and accurately reflect the ability of the ML model to detect attacks is imperative, as the IDS is responsible for protecting the essential information system against security threats. However, implementing an ML model with an imbalanced dataset presents challenges such as biased results, overfitting, and difficulties in evaluating performance using traditional metrics. When developing the ML model, it is necessary to handle the imbalanced nature of the dataset carefully.

A real-time detection system in the IoMT using ML is crucial for prompt and adequate decision-making in the event of security threats. However, implementing such a system presents several technical challenges, including the efficient and accurate processing of a large volume of data generated by the IoMT system, ensuring a high level of accuracy in the ML model's detection of attacks, and low latency in processing and prediction. The system must also be scalable and have the hardware capabilities to support the ML model and handle large amounts of data in real-time.

The use of black-box models in applications like healthcare is limited due to regulations, such as the General Data Protection Regulation, which prohibits automated decisions in critical sectors. Despite their widespread use in other applications, black-box models present significant challenges, including a lack of transparency, making it difficult to understand how the model arrived at its predictions and identify potential biases. Explaining the model's decisions is also a significant challenge, especially in applications where justifiable decisions are indispensable. Black-box models can also be vulnerable to adversarial attacks and have limited interpretability, making it challenging to debug and fine-tune the model to improve its performance. Explainable artificial intelligence (XAI) has become increasingly popular in addressing these challenges. XAI allows for creation of models that can be interpreted, understood, and transparent, and the reasoning behind their decisions can be easily explained. By using XAI, the limitations of black-box models can be overcome, promoting responsible AI development and deployment, improving trust in the model, and increasing accountability.

Implementing a real-time detection system for the IoMT using ML requires using a dataset that accurately represents the IoMT system. The utilization of datasets not designed explicitly for the IoMT system may result in a limited representation of the system and its potential attack scenarios, hindering the ability of the ML model to detect attacks effectively in real-world situations. Obtaining a comprehensive dataset encompassing the diverse range of attacks and communication protocols within the IoMT system remains challenging.

The IoMT system requires effective IDS to ensure the safety and security of medical data. Various architectures have been proposed to support these requirements, including centralized, distributed, fog-cloud and federated solutions. While centralized solutions, such as SDN, offer a centralized control point, they also pose challenges, such as the risk of a single point of failure, increased latency, and privacy concerns. Distributed solutions present a unique approach, but the practical deployment of these solutions in real-world situations still needs to be improved \cite{thamilarasu2020intrusion}. The use of a fog-cloud architecture presents a promising approach. However, the challenge of submerged data flow must be addressed in this architecture. On the other hand, federated solutions must consider the risk of adversarial attacks during the model-sharing process.

\subsection{Storage Level Security}

The concept of IDS in the medical server encounters several challenges and limitations. The data structure and format utilized for training the ML model within the medical server are distinct from other servers, making it challenging to extend the model's applicability to other medical institutions. 

A large amount of unlabeled data for supervised learning processes presents a significant obstacle for ML practitioners. The lack of labeled data makes it challenging to train the model effectively. The manual data labeling process, which often requires an interview with the patient to ensure correct labeling, can be both time-consuming and resource-intensive. Notably, this manual labeling process and the need for patient interaction result in intrusion detection at the medical server level is not fully automated. Therefore, unsupervised or semi-supervised ML methods should be further investigated at this level with particular attention to reducing false positives.

The possibility of a malicious actor altering a single feature value within an EHR or EMR for a single patient poses a challenge for ML models trained on a large amount of medical data \cite{mcglade2019ml}. These models typically rely on identifying patterns within the data and are designed to detect larger deviations from these patterns, making it difficult for the models to detect a single, subtle change in a feature value. Such an attack could have significant consequences, particularly in a medical context where patient data integrity is vital. Measures must be taken to address this challenge, such as utilizing more sophisticated ML models that can detect single-value changes or augmenting the training data with a diverse range of feature changes to enhance the model's ability to identify such modifications.

The difficulty in obtaining public medical data in the form of EMRs or EHRs results from such data's sensitivity and confidentiality. The privacy and security of patients and medical information are of utmost importance. Therefore, the release of such data is often heavily regulated and subject to strict privacy laws and regulations. This presents a significant challenge for those seeking to use the data for research or analysis purposes, as access to a large and diverse sample of medical data is critical for developing and training ML models. 

\section{CONCLUSION} \label{CONCLUSION}
A comprehensive survey on how to use an IDS-based ML to secure the IoMT is conducted. For this purpose, the generic architecture of IoMT, which is divided into three layers (data acquisition layer, personal server layer, and medical server layer) is presented. Then the requirements and possible threats that can affect the security of IoMT are provided. Next, the ML-based solutions for IoMT security are reviewed and categorized into three levels: data collection level, transmission level, and storage level, indicating the advantages, disadvantages, and datasets used. Finally, the challenges and limitations of using ML in these categories are discussed. This survey aims to highlight ML ability to bring security to complex infrastructures such as IoMT and the capacity to comply with the particular constraints of IoMT.

The main limitations of this survey are as follows: (i) The literature survey that is the subject of this study may have unintentionally left out the most current non-scientific developments or academic works that were not yet published at the time of the investigation. (ii) Although every attempt has been made to provide background, this research includes several concepts that may need to be referred to additional sources for a complete comprehension. (iii) It is noteworthy that our recommendations are inherently subjective. We intend to encourage additional research and discussion to address these limitations.

\section{CRediT authorship contribution statement}
\textbf{Ayoub Si-ahmed:} Methodology, Investigation, Writing - Original Draft
\textbf{Mohammed Ali Al-Garadi:} Methodology, Writing - Review \& Editing, Supervision, Project administration.
\textbf{Narhimene Boustia:} Resources, Writing - Review \& Editing, Supervision, Project administration.

\section{Declaration of Competing Interest}
The authors declare that they have no known competing financial interests or personal relationships that could have appeared to influence the work reported in this paper.

\section{Acknowledgments}
This study was financed by the laboratory LRDSI of the university Blida 1, Algeria.
\bibliographystyle{model1-num-names}

\bibliography{cas-refs}

\begin{thebibliography}{114}
\expandafter\ifx\csname natexlab\endcsname\relax\def\natexlab#1{#1}\fi
\providecommand{\url}[1]{\texttt{#1}}
\providecommand{\href}[2]{#2}
\providecommand{\path}[1]{#1}
\providecommand{\DOIprefix}{doi:}
\providecommand{\ArXivprefix}{arXiv:}
\providecommand{\URLprefix}{URL: }
\providecommand{\Pubmedprefix}{pmid:}
\providecommand{\doi}[1]{\href{http://dx.doi.org/#1}{\path{#1}}}
\providecommand{\Pubmed}[1]{\href{pmid:#1}{\path{#1}}}
\providecommand{\bibinfo}[2]{#2}
\ifx\xfnm\relax \def\xfnm[#1]{\unskip,\space#1}\fi
\bibitem[{Suresh et~al.(2014)Suresh, Daniel, Parthasarathy, and
  Aswathy}]{suresh2014state}
\bibinfo{author}{P.~Suresh}, \bibinfo{author}{J.~V. Daniel},
  \bibinfo{author}{V.~Parthasarathy}, \bibinfo{author}{R.~Aswathy},
\newblock \bibinfo{title}{A state of the art review on the internet of things
  (iot) history, technology and fields of deployment},
\newblock in: \bibinfo{booktitle}{2014 International conference on science
  engineering and management research (ICSEMR)}, \bibinfo{organization}{IEEE},
  \bibinfo{year}{2014}, pp. \bibinfo{pages}{1--8}.
\bibitem[{wik(2022)}]{wikipedia_2022}
\bibinfo{title}{Internet of things}, \bibinfo{year}{2022}. \URLprefix
  \url{https://en.wikipedia.org/wiki/Internet_of_things},
  \bibinfo{note}{(Accessed Feb. 04, 2022)}.
\bibitem[{tol(2022)}]{tollefson_tollefson}
\bibinfo{title}{Global iot market to grow to 24.1 billion devices in 2030,
  generating \$1.5 trillion annual revenue}, \bibinfo{year}{2022}. \URLprefix
  \url{https://transformainsights.com/news/iot-market-24-billion-usd15-trillion-revenue-2030},
  \bibinfo{note}{(Accessed Feb. 04, 2022)}.
\bibitem[{Sie(2022)}]{SiemensHealthcare2019}
\bibinfo{title}{Embracing healthcare 4.0}, \bibinfo{year}{2022}. \URLprefix
  \url{https://www.siemens-healthineers.com/insights/news/embracing-healthcare-4-0.html},
  \bibinfo{note}{(Accessed Feb. 04, 2022)}.
\bibitem[{Rathore et~al.(2019)Rathore, Al-Ali, Mohamed, Du, and
  Guizani}]{rathore2019novel}
\bibinfo{author}{H.~Rathore}, \bibinfo{author}{A.~K. Al-Ali},
  \bibinfo{author}{A.~Mohamed}, \bibinfo{author}{X.~Du},
  \bibinfo{author}{M.~Guizani},
\newblock \bibinfo{title}{A novel deep learning strategy for classifying
  different attack patterns for deep brain implants},
\newblock \bibinfo{journal}{IEEE Access} \bibinfo{volume}{7}
  (\bibinfo{year}{2019}) \bibinfo{pages}{24154--24164}.
\bibitem[{Rathore et~al.(2017)Rathore, Al-Ali, Mohamed, Du, and
  Guizani}]{rathore2017dlrt}
\bibinfo{author}{H.~Rathore}, \bibinfo{author}{A.~Al-Ali},
  \bibinfo{author}{A.~Mohamed}, \bibinfo{author}{X.~Du},
  \bibinfo{author}{M.~Guizani},
\newblock \bibinfo{title}{Dlrt: Deep learning approach for reliable diabetic
  treatment},
\newblock in: \bibinfo{booktitle}{GLOBECOM 2017-2017 IEEE global communications
  conference}, \bibinfo{organization}{IEEE}, \bibinfo{year}{2017}, pp.
  \bibinfo{pages}{1--6}.
\bibitem[{Kintzlinger et~al.(2020)Kintzlinger, Cohen, Nissim, Rav-Acha,
  Khalameizer, Elovici, Shahar, and Katz}]{kintzlinger2020cardiwall}
\bibinfo{author}{M.~Kintzlinger}, \bibinfo{author}{A.~Cohen},
  \bibinfo{author}{N.~Nissim}, \bibinfo{author}{M.~Rav-Acha},
  \bibinfo{author}{V.~Khalameizer}, \bibinfo{author}{Y.~Elovici},
  \bibinfo{author}{Y.~Shahar}, \bibinfo{author}{A.~Katz},
\newblock \bibinfo{title}{Cardiwall: a trusted firewall for the detection of
  malicious clinical programming of cardiac implantable electronic devices},
\newblock \bibinfo{journal}{IEEE Access} \bibinfo{volume}{8}
  (\bibinfo{year}{2020}) \bibinfo{pages}{48123--48140}.
\bibitem[{McLaughlin et~al.(2017)McLaughlin, Martinez~del Rincon, Kang, Yerima,
  Miller, Sezer, Safaei, Trickel, Zhao, Doup{\'e} et~al.}]{mclaughlin2017deep}
\bibinfo{author}{N.~McLaughlin}, \bibinfo{author}{J.~Martinez~del Rincon},
  \bibinfo{author}{B.~Kang}, \bibinfo{author}{S.~Yerima},
  \bibinfo{author}{P.~Miller}, \bibinfo{author}{S.~Sezer},
  \bibinfo{author}{Y.~Safaei}, \bibinfo{author}{E.~Trickel},
  \bibinfo{author}{Z.~Zhao}, \bibinfo{author}{A.~Doup{\'e}}, et~al.,
\newblock \bibinfo{title}{Deep android malware detection},
\newblock in: \bibinfo{booktitle}{Proceedings of the seventh ACM on conference
  on data and application security and privacy}, \bibinfo{year}{2017}, pp.
  \bibinfo{pages}{301--308}.
\bibitem[{Rathore et~al.(2017)Rathore, Mohamed, Al-Ali, Du, and
  Guizani}]{rathore2017review}
\bibinfo{author}{H.~Rathore}, \bibinfo{author}{A.~Mohamed},
  \bibinfo{author}{A.~Al-Ali}, \bibinfo{author}{X.~Du},
  \bibinfo{author}{M.~Guizani},
\newblock \bibinfo{title}{A review of security challenges, attacks and
  resolutions for wireless medical devices},
\newblock in: \bibinfo{booktitle}{2017 13th International Wireless
  Communications and Mobile Computing Conference (IWCMC)},
  \bibinfo{organization}{IEEE}, \bibinfo{year}{2017}, pp.
  \bibinfo{pages}{1495--1501}.
\bibitem[{Hussain et~al.(2019)Hussain, Mehmood, Khan, Khan, and
  Iqbal}]{hussain2019authentication}
\bibinfo{author}{M.~Hussain}, \bibinfo{author}{A.~Mehmood},
  \bibinfo{author}{S.~Khan}, \bibinfo{author}{M.~A. Khan},
  \bibinfo{author}{Z.~Iqbal},
\newblock \bibinfo{title}{Authentication techniques and methodologies used in
  wireless body area networks},
\newblock \bibinfo{journal}{Journal of Systems Architecture}
  \bibinfo{volume}{101} (\bibinfo{year}{2019}) \bibinfo{pages}{101655}.
\bibitem[{Wazid et~al.(2019)Wazid, Das, Rodrigues, Shetty, and
  Park}]{wazid2019iomt}
\bibinfo{author}{M.~Wazid}, \bibinfo{author}{A.~K. Das}, \bibinfo{author}{J.~J.
  Rodrigues}, \bibinfo{author}{S.~Shetty}, \bibinfo{author}{Y.~Park},
\newblock \bibinfo{title}{Iomt malware detection approaches: analysis and
  research challenges},
\newblock \bibinfo{journal}{IEEE Access} \bibinfo{volume}{7}
  (\bibinfo{year}{2019}) \bibinfo{pages}{182459--182476}.
\bibitem[{Newaz et~al.(2021)Newaz, Sikder, Rahman, and
  Uluagac}]{newaz2021survey}
\bibinfo{author}{A.~I. Newaz}, \bibinfo{author}{A.~K. Sikder},
  \bibinfo{author}{M.~A. Rahman}, \bibinfo{author}{A.~S. Uluagac},
\newblock \bibinfo{title}{A survey on security and privacy issues in modern
  healthcare systems: Attacks and defenses},
\newblock \bibinfo{journal}{ACM Transactions on Computing for Healthcare}
  \bibinfo{volume}{2} (\bibinfo{year}{2021}) \bibinfo{pages}{1--44}.
\bibitem[{Narwal and Mohapatra(2021)}]{narwal2021survey}
\bibinfo{author}{B.~Narwal}, \bibinfo{author}{A.~K. Mohapatra},
\newblock \bibinfo{title}{A survey on security and authentication in wireless
  body area networks},
\newblock \bibinfo{journal}{Journal of Systems Architecture}
  \bibinfo{volume}{113} (\bibinfo{year}{2021}) \bibinfo{pages}{101883}.
\bibitem[{Saxena and Mittal(2022)}]{saxena2022internet}
\bibinfo{author}{A.~Saxena}, \bibinfo{author}{S.~Mittal},
\newblock \bibinfo{title}{Internet of medical things (iomt) security and
  privacy: A survey of recent advances and enabling technologies},
\newblock in: \bibinfo{booktitle}{Proceedings of the 2022 Fourteenth
  International Conference on Contemporary Computing}, \bibinfo{year}{2022},
  pp. \bibinfo{pages}{550--559}.
\bibitem[{Hameed et~al.(2021)Hameed, Hassan, Latiff, and
  Ghabban}]{hameed2021systematic}
\bibinfo{author}{S.~S. Hameed}, \bibinfo{author}{W.~H. Hassan},
  \bibinfo{author}{L.~A. Latiff}, \bibinfo{author}{F.~Ghabban},
\newblock \bibinfo{title}{A systematic review of security and privacy issues in
  the internet of medical things; the role of machine learning approaches},
\newblock \bibinfo{journal}{PeerJ Computer Science} \bibinfo{volume}{7}
  (\bibinfo{year}{2021}) \bibinfo{pages}{e414}.
\bibitem[{Rahmani et~al.(2018)Rahmani, Gia, Negash, Anzanpour, Azimi, Jiang,
  and Liljeberg}]{rahmani2018exploiting}
\bibinfo{author}{A.~M. Rahmani}, \bibinfo{author}{T.~N. Gia},
  \bibinfo{author}{B.~Negash}, \bibinfo{author}{A.~Anzanpour},
  \bibinfo{author}{I.~Azimi}, \bibinfo{author}{M.~Jiang},
  \bibinfo{author}{P.~Liljeberg},
\newblock \bibinfo{title}{Exploiting smart e-health gateways at the edge of
  healthcare internet-of-things: A fog computing approach},
\newblock \bibinfo{journal}{Future Generation Computer Systems}
  \bibinfo{volume}{78} (\bibinfo{year}{2018}) \bibinfo{pages}{641--658}.
\bibitem[{Irfan and Ahmad(2018)}]{irfan2018internet}
\bibinfo{author}{M.~Irfan}, \bibinfo{author}{N.~Ahmad},
\newblock \bibinfo{title}{Internet of medical things: Architectural model,
  motivational factors and impediments},
\newblock in: \bibinfo{booktitle}{2018 15th learning and technology conference
  (L\&T)}, \bibinfo{organization}{IEEE}, \bibinfo{year}{2018}, pp.
  \bibinfo{pages}{6--13}.
\bibitem[{Khan et~al.(2012)Khan, Khan, Zaheer, and Khan}]{khan2012future}
\bibinfo{author}{R.~Khan}, \bibinfo{author}{S.~U. Khan},
  \bibinfo{author}{R.~Zaheer}, \bibinfo{author}{S.~Khan},
\newblock \bibinfo{title}{Future internet: the internet of things architecture,
  possible applications and key challenges},
\newblock in: \bibinfo{booktitle}{2012 10th international conference on
  frontiers of information technology}, \bibinfo{organization}{IEEE},
  \bibinfo{year}{2012}, pp. \bibinfo{pages}{257--260}.
\bibitem[{Kumar et~al.(2021)Kumar, Gupta, and Tripathi}]{kumar2021ensemble}
\bibinfo{author}{P.~Kumar}, \bibinfo{author}{G.~P. Gupta},
  \bibinfo{author}{R.~Tripathi},
\newblock \bibinfo{title}{An ensemble learning and fog-cloud
  architecture-driven cyber-attack detection framework for iomt networks},
\newblock \bibinfo{journal}{Computer Communications} \bibinfo{volume}{166}
  (\bibinfo{year}{2021}) \bibinfo{pages}{110--124}.
\bibitem[{Khan and Akhunzada(2021)}]{khan2021hybrid}
\bibinfo{author}{S.~Khan}, \bibinfo{author}{A.~Akhunzada},
\newblock \bibinfo{title}{A hybrid dl-driven intelligent sdn-enabled malware
  detection framework for internet of medical things (iomt)},
\newblock \bibinfo{journal}{Computer Communications} \bibinfo{volume}{170}
  (\bibinfo{year}{2021}) \bibinfo{pages}{209--216}.
\bibitem[{Chakraborty et~al.(2019)Chakraborty, Aich, and
  Kim}]{chakraborty2019secure}
\bibinfo{author}{S.~Chakraborty}, \bibinfo{author}{S.~Aich},
  \bibinfo{author}{H.-C. Kim},
\newblock \bibinfo{title}{A secure healthcare system design framework using
  blockchain technology},
\newblock in: \bibinfo{booktitle}{2019 21st International Conference on
  Advanced Communication Technology (ICACT)}, \bibinfo{organization}{IEEE},
  \bibinfo{year}{2019}, pp. \bibinfo{pages}{260--264}.
\bibitem[{Moosavi et~al.(2015)Moosavi, Gia, Rahmani, Nigussie, Virtanen,
  Isoaho, and Tenhunen}]{moosavi2015sea}
\bibinfo{author}{S.~R. Moosavi}, \bibinfo{author}{T.~N. Gia},
  \bibinfo{author}{A.-M. Rahmani}, \bibinfo{author}{E.~Nigussie},
  \bibinfo{author}{S.~Virtanen}, \bibinfo{author}{J.~Isoaho},
  \bibinfo{author}{H.~Tenhunen},
\newblock \bibinfo{title}{Sea: a secure and efficient authentication and
  authorization architecture for iot-based healthcare using smart gateways},
\newblock \bibinfo{journal}{Procedia Computer Science} \bibinfo{volume}{52}
  (\bibinfo{year}{2015}) \bibinfo{pages}{452--459}.
\bibitem[{Zhang et~al.(2015)Zhang, Yang, Liang, Su, Shen, and
  Luo}]{zhang2015security}
\bibinfo{author}{K.~Zhang}, \bibinfo{author}{K.~Yang},
  \bibinfo{author}{X.~Liang}, \bibinfo{author}{Z.~Su},
  \bibinfo{author}{X.~Shen}, \bibinfo{author}{H.~H. Luo},
\newblock \bibinfo{title}{Security and privacy for mobile healthcare networks:
  from a quality of protection perspective},
\newblock \bibinfo{journal}{IEEE Wireless Communications} \bibinfo{volume}{22}
  (\bibinfo{year}{2015}) \bibinfo{pages}{104--112}.
\bibitem[{Sun et~al.(2019)Sun, Lo, and Lo}]{sun2019security}
\bibinfo{author}{Y.~Sun}, \bibinfo{author}{F.~P.-W. Lo},
  \bibinfo{author}{B.~Lo},
\newblock \bibinfo{title}{Security and privacy for the internet of medical
  things enabled healthcare systems: A survey},
\newblock \bibinfo{journal}{IEEE Access} \bibinfo{volume}{7}
  (\bibinfo{year}{2019}) \bibinfo{pages}{183339--183355}.
\bibitem[{Arya and Gore(2020)}]{arya2020data}
\bibinfo{author}{K.~Arya}, \bibinfo{author}{R.~Gore},
\newblock \bibinfo{title}{Data security for wban in e-health iot applications},
\newblock in: \bibinfo{booktitle}{Intelligent Data Security Solutions for
  e-Health Applications}, \bibinfo{publisher}{Elsevier}, \bibinfo{year}{2020},
  pp. \bibinfo{pages}{205--218}.
\bibitem[{Hathaliya and Tanwar(2020)}]{hathaliya2020exhaustive}
\bibinfo{author}{J.~J. Hathaliya}, \bibinfo{author}{S.~Tanwar},
\newblock \bibinfo{title}{An exhaustive survey on security and privacy issues
  in healthcare 4.0},
\newblock \bibinfo{journal}{Computer Communications} \bibinfo{volume}{153}
  (\bibinfo{year}{2020}) \bibinfo{pages}{311--335}.
\bibitem[{Alkeem et~al.(2017)Alkeem, Shehada, Yeun, Zemerly, and
  Hu}]{alkeem2017new}
\bibinfo{author}{E.~A. Alkeem}, \bibinfo{author}{D.~Shehada},
  \bibinfo{author}{C.~Y. Yeun}, \bibinfo{author}{M.~J. Zemerly},
  \bibinfo{author}{J.~Hu},
\newblock \bibinfo{title}{New secure healthcare system using cloud of things},
\newblock \bibinfo{journal}{Cluster Computing} \bibinfo{volume}{20}
  (\bibinfo{year}{2017}) \bibinfo{pages}{2211--2229}.
\bibitem[{Roy et~al.(2020)Roy, Chowdhury, and Aslam}]{roy2020security}
\bibinfo{author}{M.~Roy}, \bibinfo{author}{C.~Chowdhury},
  \bibinfo{author}{N.~Aslam},
\newblock \bibinfo{title}{Security and privacy issues in wireless sensor and
  body area networks},
\newblock in: \bibinfo{booktitle}{Handbook of computer networks and cyber
  security}, \bibinfo{publisher}{Springer}, \bibinfo{year}{2020}, pp.
  \bibinfo{pages}{173--200}.
\bibitem[{Pirbhulal et~al.(2019)Pirbhulal, Samuel, Wu, Sangaiah, and
  Li}]{pirbhulal2019joint}
\bibinfo{author}{S.~Pirbhulal}, \bibinfo{author}{O.~W. Samuel},
  \bibinfo{author}{W.~Wu}, \bibinfo{author}{A.~K. Sangaiah},
  \bibinfo{author}{G.~Li},
\newblock \bibinfo{title}{A joint resource-aware and medical data security
  framework for wearable healthcare systems},
\newblock \bibinfo{journal}{Future Generation Computer Systems}
  \bibinfo{volume}{95} (\bibinfo{year}{2019}) \bibinfo{pages}{382--391}.
\bibitem[{Kompara and H{\"o}lbl(2018)}]{kompara2018survey}
\bibinfo{author}{M.~Kompara}, \bibinfo{author}{M.~H{\"o}lbl},
\newblock \bibinfo{title}{Survey on security in intra-body area network
  communication},
\newblock \bibinfo{journal}{Ad Hoc Networks} \bibinfo{volume}{70}
  (\bibinfo{year}{2018}) \bibinfo{pages}{23--43}.
\bibitem[{Malan et~al.(2004)Malan, Fulford-Jones, Welsh, and
  Moulton}]{malan2004codeblue}
\bibinfo{author}{D.~J. Malan}, \bibinfo{author}{T.~Fulford-Jones},
  \bibinfo{author}{M.~Welsh}, \bibinfo{author}{S.~Moulton},
\newblock \bibinfo{title}{Codeblue: An ad hoc sensor network infrastructure for
  emergency medical care},
\newblock in: \bibinfo{booktitle}{International workshop on wearable and
  implantable body sensor networks}, \bibinfo{year}{2004}.
\bibitem[{Ko et~al.(2010)Ko, Lim, Chen, Musvaloiu-E, Terzis, Masson, Gao,
  Destler, Selavo, and Dutton}]{ko2010medisn}
\bibinfo{author}{J.~Ko}, \bibinfo{author}{J.~H. Lim},
  \bibinfo{author}{Y.~Chen}, \bibinfo{author}{R.~Musvaloiu-E},
  \bibinfo{author}{A.~Terzis}, \bibinfo{author}{G.~M. Masson},
  \bibinfo{author}{T.~Gao}, \bibinfo{author}{W.~Destler},
  \bibinfo{author}{L.~Selavo}, \bibinfo{author}{R.~P. Dutton},
\newblock \bibinfo{title}{Medisn: Medical emergency detection in sensor
  networks},
\newblock \bibinfo{journal}{ACM Transactions on Embedded Computing Systems
  (TECS)} \bibinfo{volume}{10} (\bibinfo{year}{2010}) \bibinfo{pages}{1--29}.
\bibitem[{Segovia et~al.(2013)Segovia, Gramp{\'\i}n, and
  Baliosian}]{segovia2013analysis}
\bibinfo{author}{M.~Segovia}, \bibinfo{author}{E.~Gramp{\'\i}n},
  \bibinfo{author}{J.~Baliosian},
\newblock \bibinfo{title}{Analysis of the applicability of wireless sensor
  networks attacks to body area networks},
\newblock in: \bibinfo{booktitle}{Proceedings of the 8th International
  Conference on Body Area Networks}, \bibinfo{year}{2013}, pp.
  \bibinfo{pages}{509--512}.
\bibitem[{Kintzlinger and Nissim(2019)}]{kintzlinger2019keep}
\bibinfo{author}{M.~Kintzlinger}, \bibinfo{author}{N.~Nissim},
\newblock \bibinfo{title}{Keep an eye on your personal belongings! the security
  of personal medical devices and their ecosystems},
\newblock \bibinfo{journal}{Journal of biomedical informatics}
  \bibinfo{volume}{95} (\bibinfo{year}{2019}) \bibinfo{pages}{103233}.
\bibitem[{Bangash et~al.(2017)Bangash, Al-Salhi et~al.}]{bangash2017security}
\bibinfo{author}{Y.~A. Bangash}, \bibinfo{author}{Y.~E. Al-Salhi}, et~al.,
\newblock \bibinfo{title}{Security issues and challenges in wireless sensor
  networks: A survey.},
\newblock \bibinfo{journal}{IAENG International Journal of Computer Science}
  \bibinfo{volume}{44} (\bibinfo{year}{2017}).
\bibitem[{Hei et~al.(2014)Hei, Du, Lin, Lee, and Sokolsky}]{hei2014patient}
\bibinfo{author}{X.~Hei}, \bibinfo{author}{X.~Du}, \bibinfo{author}{S.~Lin},
  \bibinfo{author}{I.~Lee}, \bibinfo{author}{O.~Sokolsky},
\newblock \bibinfo{title}{Patient infusion pattern based access control schemes
  for wireless insulin pump system},
\newblock \bibinfo{journal}{IEEE Transactions on Parallel and Distributed
  Systems} \bibinfo{volume}{26} (\bibinfo{year}{2014})
  \bibinfo{pages}{3108--3121}.
\bibitem[{Doss et~al.(2015)Doss, Piramuthu, and Wei}]{doss2015future}
\bibinfo{author}{R.~Doss}, \bibinfo{author}{S.~Piramuthu},
  \bibinfo{author}{Z.~Wei}, \bibinfo{title}{Future Network Systems and
  Security: First International Conference, FNSS 2015, Paris, France, June
  11-13, 2015, Proceedings}, volume \bibinfo{volume}{523},
  \bibinfo{publisher}{Springer}, \bibinfo{year}{2015}.
\bibitem[{Rani and Kumar(2017)}]{rani2017survey}
\bibinfo{author}{A.~Rani}, \bibinfo{author}{S.~Kumar},
\newblock \bibinfo{title}{A survey of security in wireless sensor networks},
\newblock in: \bibinfo{booktitle}{2017 3rd International Conference on
  Computational Intelligence \& Communication Technology (CICT)},
  \bibinfo{organization}{IEEE}, \bibinfo{year}{2017}, pp.
  \bibinfo{pages}{1--5}.
\bibitem[{Xu et~al.(2016)Xu, Wu, Daneshmand, Liu, and Wang}]{xu2016data}
\bibinfo{author}{G.~Xu}, \bibinfo{author}{Q.~Wu},
  \bibinfo{author}{M.~Daneshmand}, \bibinfo{author}{Y.~Liu},
  \bibinfo{author}{M.~Wang},
\newblock \bibinfo{title}{A data privacy protective mechanism for wireless body
  area networks},
\newblock \bibinfo{journal}{wireless communications and mobile computing}
  \bibinfo{volume}{16} (\bibinfo{year}{2016}) \bibinfo{pages}{1746--1758}.
\bibitem[{Sanei et~al.(2020)Sanei, Jarchi, and Constantinides}]{Sanei2020}
\bibinfo{author}{S.~Sanei}, \bibinfo{author}{D.~Jarchi}, \bibinfo{author}{A.~G.
  Constantinides},
\newblock \bibinfo{title}{{Quality of Service, Security, and Privacy for
  Wearable Sensor Data}},
\newblock \bibinfo{journal}{Body Sensor Networking, Design and Algorithms}
  (\bibinfo{year}{2020}) \bibinfo{pages}{325--343}.
\bibitem[{Habib et~al.(2015)Habib, Torjusen, and Leister}]{habib2015security}
\bibinfo{author}{K.~Habib}, \bibinfo{author}{A.~Torjusen},
  \bibinfo{author}{W.~Leister},
\newblock \bibinfo{title}{Security analysis of a patient monitoring system for
  the internet of things in ehealth},
\newblock in: \bibinfo{booktitle}{The Seventh International Conference on
  eHealth, Telemedicine, and Social Medicine (eTELEMED)}, volume
  \bibinfo{volume}{335}, \bibinfo{year}{2015}.
\bibitem[{Maheswar et~al.(2019)Maheswar, Kanagachidambaresan, Jayaparvathy, and
  Thampi}]{maheswar2019body}
\bibinfo{author}{R.~Maheswar}, \bibinfo{author}{G.~Kanagachidambaresan},
  \bibinfo{author}{R.~Jayaparvathy}, \bibinfo{author}{S.~M. Thampi},
  \bibinfo{title}{Body area network challenges and solutions},
  \bibinfo{publisher}{Springer}, \bibinfo{year}{2019}.
\bibitem[{Kumar and Lee(2012)}]{kumar2012security}
\bibinfo{author}{P.~Kumar}, \bibinfo{author}{H.-J. Lee},
\newblock \bibinfo{title}{Security issues in healthcare applications using
  wireless medical sensor networks: A survey},
\newblock \bibinfo{journal}{sensors} \bibinfo{volume}{12}
  (\bibinfo{year}{2012}) \bibinfo{pages}{55--91}.
\bibitem[{Pathania and Bilandi(2014)}]{pathania2014security}
\bibinfo{author}{S.~Pathania}, \bibinfo{author}{N.~Bilandi},
\newblock \bibinfo{title}{Security issues in wireless body area network},
\newblock \bibinfo{journal}{Int J Comput Sci Mobile Comput} \bibinfo{volume}{3}
  (\bibinfo{year}{2014}) \bibinfo{pages}{1171--1178}.
\bibitem[{Stavroulakis and Stamp(2010)}]{stavroulakis2010handbook}
\bibinfo{author}{P.~Stavroulakis}, \bibinfo{author}{M.~Stamp},
  \bibinfo{title}{Handbook of information and communication security},
  \bibinfo{publisher}{Springer Science \& Business Media},
  \bibinfo{year}{2010}.
\bibitem[{Hamid et~al.(2006)Hamid, Rashid, and Hong}]{hamid2006routing}
\bibinfo{author}{M.~A. Hamid}, \bibinfo{author}{M.~Rashid},
  \bibinfo{author}{C.~S. Hong},
\newblock \bibinfo{title}{Routing security in sensor network: Hello flood
  attack and defense},
\newblock \bibinfo{journal}{IEEE ICNEWS} \bibinfo{volume}{2}
  (\bibinfo{year}{2006}) \bibinfo{pages}{2--4}.
\bibitem[{Tumrongwittayapak and
  Varakulsiripunth(2009)}]{tumrongwittayapak2009detecting}
\bibinfo{author}{C.~Tumrongwittayapak}, \bibinfo{author}{R.~Varakulsiripunth},
\newblock \bibinfo{title}{Detecting sinkhole attack and selective forwarding
  attack in wireless sensor networks},
\newblock in: \bibinfo{booktitle}{2009 7th International Conference on
  Information, Communications and Signal Processing (ICICS)},
  \bibinfo{organization}{IEEE}, \bibinfo{year}{2009}, pp.
  \bibinfo{pages}{1--5}.
\bibitem[{Mana et~al.(2009)Mana, Feham, and Bensaber}]{mana2009sekeban}
\bibinfo{author}{M.~Mana}, \bibinfo{author}{M.~Feham}, \bibinfo{author}{B.~A.
  Bensaber},
\newblock \bibinfo{title}{Sekeban (secure and efficient key exchange for
  wireless body},
\newblock in: \bibinfo{booktitle}{Area Network)”, International Journal of
  Advanced Science and Technology}, \bibinfo{organization}{Citeseer},
  \bibinfo{year}{2009}.
\bibitem[{Stavros(2002)}]{stavros2002advances}
\bibinfo{author}{A.~V. Stavros}, \bibinfo{title}{Advances in Communications and
  Media Research}, volume~\bibinfo{volume}{2}, \bibinfo{publisher}{Nova
  Publishers}, \bibinfo{year}{2002}.
\bibitem[{Siva~Bharathi and Venkateswari(2019)}]{siva2019security}
\bibinfo{author}{K.~Siva~Bharathi}, \bibinfo{author}{R.~Venkateswari},
\newblock \bibinfo{title}{Security challenges and solutions for wireless body
  area networks},
\newblock in: \bibinfo{booktitle}{Computing, Communication and Signal
  Processing}, \bibinfo{publisher}{Springer}, \bibinfo{year}{2019}, pp.
  \bibinfo{pages}{275--283}.
\bibitem[{Gupta et~al.(2020)Gupta, Perez, Agrawal, and
  Gupta}]{gupta2020handbook}
\bibinfo{author}{B.~B. Gupta}, \bibinfo{author}{G.~M. Perez},
  \bibinfo{author}{D.~P. Agrawal}, \bibinfo{author}{D.~Gupta},
\newblock \bibinfo{title}{Handbook of computer networks and cyber security},
\newblock \bibinfo{journal}{Springer} \bibinfo{volume}{10}
  (\bibinfo{year}{2020}) \bibinfo{pages}{978--3}.
\bibitem[{Masdari and Ahmadzadeh(2016)}]{masdari2016comprehensive}
\bibinfo{author}{M.~Masdari}, \bibinfo{author}{S.~Ahmadzadeh},
\newblock \bibinfo{title}{Comprehensive analysis of the authentication methods
  in wireless body area networks},
\newblock \bibinfo{journal}{Security and communication networks}
  \bibinfo{volume}{9} (\bibinfo{year}{2016}) \bibinfo{pages}{4777--4803}.
\bibitem[{Niksaz and Branch(2015)}]{niksaz2015wireless}
\bibinfo{author}{P.~Niksaz}, \bibinfo{author}{M.~Branch},
\newblock \bibinfo{title}{Wireless body area networks: attacks and
  countermeasures},
\newblock \bibinfo{journal}{Int. J. Sci. Eng. Res} \bibinfo{volume}{6}
  (\bibinfo{year}{2015}) \bibinfo{pages}{556--568}.
\bibitem[{Newsome et~al.(2004)Newsome, Shi, Song, and
  Perrig}]{newsome2004sybil}
\bibinfo{author}{J.~Newsome}, \bibinfo{author}{E.~Shi},
  \bibinfo{author}{D.~Song}, \bibinfo{author}{A.~Perrig},
\newblock \bibinfo{title}{The sybil attack in sensor networks: analysis \&
  defenses},
\newblock in: \bibinfo{booktitle}{Third international symposium on information
  processing in sensor networks, 2004. IPSN 2004},
  \bibinfo{organization}{IEEE}, \bibinfo{year}{2004}, pp.
  \bibinfo{pages}{259--268}.
\bibitem[{Ramli et~al.(2010)Ramli, Zakaria, and Sumari}]{ramli2010privacy}
\bibinfo{author}{R.~Ramli}, \bibinfo{author}{N.~Zakaria},
  \bibinfo{author}{P.~Sumari},
\newblock \bibinfo{title}{Privacy issues in pervasive healthcare monitoring
  system: A review},
\newblock \bibinfo{journal}{World Acad. Sci. Eng. Technol} \bibinfo{volume}{72}
  (\bibinfo{year}{2010}) \bibinfo{pages}{741--747}.
\bibitem[{Partala et~al.(2013)Partala, Ker{\"a}neny, S{\"a}rest{\"o}niemi,
  H{\"a}m{\"a}l{\"a}inen, Iinatti, J{\"a}ms{\"a}, Reponen, and
  Sepp{\"a}nen}]{partala2013security}
\bibinfo{author}{J.~Partala}, \bibinfo{author}{N.~Ker{\"a}neny},
  \bibinfo{author}{M.~S{\"a}rest{\"o}niemi},
  \bibinfo{author}{M.~H{\"a}m{\"a}l{\"a}inen}, \bibinfo{author}{J.~Iinatti},
  \bibinfo{author}{T.~J{\"a}ms{\"a}}, \bibinfo{author}{J.~Reponen},
  \bibinfo{author}{T.~Sepp{\"a}nen},
\newblock \bibinfo{title}{Security threats against the transmission chain of a
  medical health monitoring system},
\newblock in: \bibinfo{booktitle}{2013 IEEE 15th International Conference on
  e-Health Networking, Applications and Services (Healthcom 2013)},
  \bibinfo{organization}{IEEE}, \bibinfo{year}{2013}, pp.
  \bibinfo{pages}{243--248}.
\bibitem[{Curtis et~al.(2008)Curtis, Pino, Bailey, Shih, Waterman, Vinterbo,
  Stair, Guttag, Greenes, and Ohno-Machado}]{curtis2008smart}
\bibinfo{author}{D.~W. Curtis}, \bibinfo{author}{E.~J. Pino},
  \bibinfo{author}{J.~M. Bailey}, \bibinfo{author}{E.~I. Shih},
  \bibinfo{author}{J.~Waterman}, \bibinfo{author}{S.~A. Vinterbo},
  \bibinfo{author}{T.~O. Stair}, \bibinfo{author}{J.~V. Guttag},
  \bibinfo{author}{R.~A. Greenes}, \bibinfo{author}{L.~Ohno-Machado},
\newblock \bibinfo{title}{Smart—an integrated wireless system for monitoring
  unattended patients},
\newblock \bibinfo{journal}{Journal of the American Medical Informatics
  Association} \bibinfo{volume}{15} (\bibinfo{year}{2008})
  \bibinfo{pages}{44--53}.
\bibitem[{Redondi et~al.(2010)Redondi, Tagliasacchi, Cesana, Borsani,
  Tarr{\'\i}o, and Salice}]{redondi2010laura}
\bibinfo{author}{A.~Redondi}, \bibinfo{author}{M.~Tagliasacchi},
  \bibinfo{author}{M.~Cesana}, \bibinfo{author}{L.~Borsani},
  \bibinfo{author}{P.~Tarr{\'\i}o}, \bibinfo{author}{F.~Salice},
\newblock \bibinfo{title}{Laura—localization and ubiquitous monitoring of
  patients for health care support},
\newblock in: \bibinfo{booktitle}{2010 IEEE 21st International Symposium on
  Personal, Indoor and Mobile Radio Communications Workshops},
  \bibinfo{organization}{IEEE}, \bibinfo{year}{2010}, pp.
  \bibinfo{pages}{218--222}.
\bibitem[{McGlade and Scott-Hayward(2019)}]{mcglade2019ml}
\bibinfo{author}{D.~McGlade}, \bibinfo{author}{S.~Scott-Hayward},
\newblock \bibinfo{title}{Ml-based cyber incident detection for electronic
  medical record (emr) systems},
\newblock \bibinfo{journal}{Smart Health} \bibinfo{volume}{12}
  (\bibinfo{year}{2019}) \bibinfo{pages}{3--23}.
\bibitem[{Alsubaei et~al.(2017)Alsubaei, Abuhussein, and
  Shiva}]{alsubaei2017security}
\bibinfo{author}{F.~Alsubaei}, \bibinfo{author}{A.~Abuhussein},
  \bibinfo{author}{S.~Shiva},
\newblock \bibinfo{title}{Security and privacy in the internet of medical
  things: taxonomy and risk assessment},
\newblock in: \bibinfo{booktitle}{2017 IEEE 42nd conference on local computer
  networks workshops (LCN workshops)}, \bibinfo{organization}{IEEE},
  \bibinfo{year}{2017}, pp. \bibinfo{pages}{112--120}.
\bibitem[{Hassija et~al.(2021)Hassija, Chamola, Bajpai, Zeadally
  et~al.}]{hassija2021security}
\bibinfo{author}{V.~Hassija}, \bibinfo{author}{V.~Chamola},
  \bibinfo{author}{B.~C. Bajpai}, \bibinfo{author}{S.~Zeadally}, et~al.,
\newblock \bibinfo{title}{Security issues in implantable medical devices: Fact
  or fiction?},
\newblock \bibinfo{journal}{Sustainable Cities and Society}
  \bibinfo{volume}{66} (\bibinfo{year}{2021}) \bibinfo{pages}{102552}.
\bibitem[{{Dana Ford}(2013)}]{DanaFord2013}
\bibinfo{author}{{Dana Ford}}, \bibinfo{title}{{Docs shielded Cheney
  defibrillator from hacks - CNN}}, \bibinfo{year}{2013}. \URLprefix
  \url{https://edition.cnn.com/2013/10/20/us/dick-cheney-gupta-interview/index.html},
  \bibinfo{note}{(Accessed Oct. 07, 2021)}.
\bibitem[{Bace and Mell(2001)}]{bace2001nist}
\bibinfo{author}{R.~Bace}, \bibinfo{author}{P.~Mell}, \bibinfo{title}{NIST
  special publication on intrusion detection systems}, \bibinfo{type}{Technical
  Report}, Booz-allen and Hamilton Inc MCLEAN VA, \bibinfo{year}{2001}.
\bibitem[{Samuel(2000)}]{samuel2000some}
\bibinfo{author}{A.~L. Samuel},
\newblock \bibinfo{title}{Some studies in machine learning using the game of
  checkers},
\newblock \bibinfo{journal}{IBM Journal of research and development}
  \bibinfo{volume}{44} (\bibinfo{year}{2000}) \bibinfo{pages}{206--226}.
\bibitem[{G{\'e}ron(2017)}]{geron2017hands}
\bibinfo{author}{A.~G{\'e}ron},
\newblock \bibinfo{title}{Hands-on machine learning with scikit-learn and
  tensorflow: Concepts},
\newblock \bibinfo{journal}{Tools, and Techniques to build intelligent systems}
   (\bibinfo{year}{2017}).
\bibitem[{Franklin(2005)}]{franklin2005elements}
\bibinfo{author}{J.~Franklin},
\newblock \bibinfo{title}{The elements of statistical learning: data mining,
  inference and prediction},
\newblock \bibinfo{journal}{The Mathematical Intelligencer}
  \bibinfo{volume}{27} (\bibinfo{year}{2005}) \bibinfo{pages}{83--85}.
\bibitem[{Pise and Kulkarni(2008)}]{pise2008survey}
\bibinfo{author}{N.~N. Pise}, \bibinfo{author}{P.~Kulkarni},
\newblock \bibinfo{title}{A survey of semi-supervised learning methods},
\newblock in: \bibinfo{booktitle}{2008 International conference on
  computational intelligence and security}, volume~\bibinfo{volume}{2},
  \bibinfo{organization}{IEEE}, \bibinfo{year}{2008}, pp.
  \bibinfo{pages}{30--34}.
\bibitem[{LeCun et~al.(2015)LeCun, Bengio, and Hinton}]{lecun2015deep}
\bibinfo{author}{Y.~LeCun}, \bibinfo{author}{Y.~Bengio},
  \bibinfo{author}{G.~Hinton},
\newblock \bibinfo{title}{Deep learning},
\newblock \bibinfo{journal}{nature} \bibinfo{volume}{521}
  (\bibinfo{year}{2015}) \bibinfo{pages}{436--444}.
\bibitem[{Khan et~al.(2017)Khan, Haldar, Ali, Iftikhar, Zia, and
  Zomaya}]{khan2017continuous}
\bibinfo{author}{F.~A. Khan}, \bibinfo{author}{N.~A.~H. Haldar},
  \bibinfo{author}{A.~Ali}, \bibinfo{author}{M.~Iftikhar},
  \bibinfo{author}{T.~A. Zia}, \bibinfo{author}{A.~Y. Zomaya},
\newblock \bibinfo{title}{A continuous change detection mechanism to identify
  anomalies in ecg signals for wban-based healthcare environments},
\newblock \bibinfo{journal}{IEEE Access} \bibinfo{volume}{5}
  (\bibinfo{year}{2017}) \bibinfo{pages}{13531--13544}.
\bibitem[{{Lugovaya T.S}(2005)}]{LugovayaT.S2005}
\bibinfo{author}{{Lugovaya T.S}}, \bibinfo{title}{{ECG-ID Database v1.0.0}},
  \bibinfo{year}{2005}. \URLprefix
  \url{https://www.physionet.org/content/ecgiddb/1.0.0/},
  \bibinfo{note}{(Accessed Oct. 07, 2021)}.
\bibitem[{Ahmad et~al.(2018)Ahmad, Song, Bilal, Saleem, and
  Ullah}]{ahmad2018securing}
\bibinfo{author}{U.~Ahmad}, \bibinfo{author}{H.~Song},
  \bibinfo{author}{A.~Bilal}, \bibinfo{author}{S.~Saleem},
  \bibinfo{author}{A.~Ullah},
\newblock \bibinfo{title}{Securing insulin pump system using deep learning and
  gesture recognition},
\newblock in: \bibinfo{booktitle}{2018 17th IEEE International Conference On
  Trust, Security And Privacy In Computing And Communications/12th IEEE
  International Conference On Big Data Science And Engineering
  (TrustCom/BigDataSE)}, \bibinfo{organization}{IEEE}, \bibinfo{year}{2018},
  pp. \bibinfo{pages}{1716--1719}.
\bibitem[{{Vincent Sigillito}(1990)}]{VincentSigillito1990}
\bibinfo{author}{{Vincent Sigillito}}, \bibinfo{title}{{Pima Indians Diabetes
  Database}}, \bibinfo{year}{1990}. \URLprefix
  \url{https://www.openml.org/d/37}, \bibinfo{note}{(Accessed Oct. 07, 2021)}.
\bibitem[{Shobana et~al.(2022)}]{shobana2022towards}
\bibinfo{author}{M.~Shobana}, et~al.,
\newblock \bibinfo{title}{Towards securing wireless insulin pump system using
  unsupervised deep learning technique},
\newblock \bibinfo{journal}{Iran Journal of Computer Science}
  (\bibinfo{year}{2022}).
\bibitem[{{Diabetes data set}(2022)}]{UMLRDDS2022}
\bibinfo{author}{{Diabetes data set}}, \bibinfo{title}{Uci machine learning
  repository: Diabetes data set}, \bibinfo{year}{2022}. \URLprefix
  \url{https://archive.ics.uci.edu/ml/datasets/diabetes},
  \bibinfo{note}{(Accessed Nov. 27, 2022)}.
\bibitem[{Goldberger et~al.(2000)Goldberger, LastName, Glass, Hausdorff,
  Ivanov, Mark, Mietus, Moody, Peng, and Stanley}]{Goldberger2000}
\bibinfo{author}{A.~Goldberger}, \bibinfo{author}{A.~L. LastName},
  \bibinfo{author}{L.~Glass}, \bibinfo{author}{J.~Hausdorff},
  \bibinfo{author}{P.~Ivanov}, \bibinfo{author}{R.~Mark},
  \bibinfo{author}{J.~Mietus}, \bibinfo{author}{G.~Moody},
  \bibinfo{author}{C.~Peng}, \bibinfo{author}{H.~Stanley},
  \bibinfo{title}{{Effect of Deep Brain Stimulation on Parkinsonian Tremor
  v1.0.0}}, \bibinfo{year}{2000}. \URLprefix
  \url{https://physionet.org/content/tremordb/1.0.0/}, \bibinfo{note}{(Accessed
  Oct. 07, 2021)}.
\bibitem[{Hei et~al.(2010)Hei, Du, Wu, and Hu}]{hei2010defending}
\bibinfo{author}{X.~Hei}, \bibinfo{author}{X.~Du}, \bibinfo{author}{J.~Wu},
  \bibinfo{author}{F.~Hu},
\newblock \bibinfo{title}{Defending resource depletion attacks on implantable
  medical devices},
\newblock in: \bibinfo{booktitle}{2010 IEEE global telecommunications
  conference GLOBECOM 2010}, \bibinfo{organization}{IEEE},
  \bibinfo{year}{2010}, pp. \bibinfo{pages}{1--5}.
\bibitem[{Newaz et~al.(2019)Newaz, Sikder, Rahman, and
  Uluagac}]{newaz2019healthguard}
\bibinfo{author}{A.~I. Newaz}, \bibinfo{author}{A.~K. Sikder},
  \bibinfo{author}{M.~A. Rahman}, \bibinfo{author}{A.~S. Uluagac},
\newblock \bibinfo{title}{Healthguard: A machine learning-based security
  framework for smart healthcare systems},
\newblock in: \bibinfo{booktitle}{2019 Sixth International Conference on Social
  Networks Analysis, Management and Security (SNAMS)},
  \bibinfo{organization}{IEEE}, \bibinfo{year}{2019}, pp.
  \bibinfo{pages}{389--396}.
\bibitem[{Salem et~al.(2020)Salem, Alsubhi, Mehaoua, and
  Boutaba}]{salem2020markov}
\bibinfo{author}{O.~Salem}, \bibinfo{author}{K.~Alsubhi},
  \bibinfo{author}{A.~Mehaoua}, \bibinfo{author}{R.~Boutaba},
\newblock \bibinfo{title}{Markov models for anomaly detection in wireless body
  area networks for secure health monitoring},
\newblock \bibinfo{journal}{IEEE Journal on Selected Areas in Communications}
  \bibinfo{volume}{39} (\bibinfo{year}{2020}) \bibinfo{pages}{526--540}.
\bibitem[{{Moody GB} and {Mark RG}(1996)}]{MoodyGB1996}
\bibinfo{author}{{Moody GB}}, \bibinfo{author}{{Mark RG}},
  \bibinfo{title}{{MIMIC Database v1.0.0}}, \bibinfo{year}{1996}. \URLprefix
  \url{https://www.physionet.org/content/mimicdb/1.0.0/},
  \bibinfo{note}{(Accessed Oct. 07, 2021)}.
\bibitem[{Yaqoob et~al.(2019)Yaqoob, Abbas, and
  Atiquzzaman}]{yaqoob2019security}
\bibinfo{author}{T.~Yaqoob}, \bibinfo{author}{H.~Abbas},
  \bibinfo{author}{M.~Atiquzzaman},
\newblock \bibinfo{title}{Security vulnerabilities, attacks, countermeasures,
  and regulations of networked medical devices—a review},
\newblock \bibinfo{journal}{IEEE Communications Surveys \& Tutorials}
  \bibinfo{volume}{21} (\bibinfo{year}{2019}) \bibinfo{pages}{3723--3768}.
\bibitem[{Gao and Thamilarasu(2017)}]{gao2017machine}
\bibinfo{author}{S.~Gao}, \bibinfo{author}{G.~Thamilarasu},
\newblock \bibinfo{title}{Machine-learning classifiers for security in
  connected medical devices},
\newblock in: \bibinfo{booktitle}{2017 26th International Conference on
  Computer Communication and Networks (ICCCN)}, \bibinfo{organization}{IEEE},
  \bibinfo{year}{2017}, pp. \bibinfo{pages}{1--5}.
\bibitem[{OMN(2021)}]{OMNeT_based_simulator}
\bibinfo{title}{{Castalia: An OMNeT-based simulator for low-power wireless
  networks such as Wireless Sensor Networks and Body Area Networks}},
  \bibinfo{year}{2021}. \URLprefix \url{https://github.com/boulis/Castalia},
  \bibinfo{note}{(Accessed Oct. 07, 2021)}.
\bibitem[{Al-Shaher et~al.(2017)Al-Shaher, Hameed, and
  {\c{T}}{\u{a}}pu{\c{s}}}]{al2017protect}
\bibinfo{author}{M.~A. Al-Shaher}, \bibinfo{author}{R.~T. Hameed},
  \bibinfo{author}{N.~{\c{T}}{\u{a}}pu{\c{s}}},
\newblock \bibinfo{title}{Protect healthcare system based on intelligent
  techniques},
\newblock in: \bibinfo{booktitle}{2017 4th International Conference on Control,
  Decision and Information Technologies (CoDIT)}, \bibinfo{organization}{IEEE},
  \bibinfo{year}{2017}, pp. \bibinfo{pages}{0421--0426}.
\bibitem[{He et~al.(2019)He, Qiao, Gao, Zheng, Chan, Li, and
  Guizani}]{he2019intrusion}
\bibinfo{author}{D.~He}, \bibinfo{author}{Q.~Qiao}, \bibinfo{author}{Y.~Gao},
  \bibinfo{author}{J.~Zheng}, \bibinfo{author}{S.~Chan},
  \bibinfo{author}{J.~Li}, \bibinfo{author}{N.~Guizani},
\newblock \bibinfo{title}{Intrusion detection based on stacked autoencoder for
  connected healthcare systems},
\newblock \bibinfo{journal}{IEEE Network} \bibinfo{volume}{33}
  (\bibinfo{year}{2019}) \bibinfo{pages}{64--69}.
\bibitem[{Newaz et~al.(2020)Newaz, Sikder, Babun, and Uluagac}]{newaz2020heka}
\bibinfo{author}{A.~I. Newaz}, \bibinfo{author}{A.~K. Sikder},
  \bibinfo{author}{L.~Babun}, \bibinfo{author}{A.~S. Uluagac},
\newblock \bibinfo{title}{Heka: A novel intrusion detection system for attacks
  to personal medical devices},
\newblock in: \bibinfo{booktitle}{2020 IEEE Conference on Communications and
  Network Security (CNS)}, \bibinfo{organization}{IEEE}, \bibinfo{year}{2020},
  pp. \bibinfo{pages}{1--9}.
\bibitem[{RM et~al.(2020)RM, Maddikunta, Parimala, Koppu, Gadekallu, Chowdhary,
  and Alazab}]{rm2020effective}
\bibinfo{author}{S.~P. RM}, \bibinfo{author}{P.~K.~R. Maddikunta},
  \bibinfo{author}{M.~Parimala}, \bibinfo{author}{S.~Koppu},
  \bibinfo{author}{T.~R. Gadekallu}, \bibinfo{author}{C.~L. Chowdhary},
  \bibinfo{author}{M.~Alazab},
\newblock \bibinfo{title}{An effective feature engineering for dnn using hybrid
  pca-gwo for intrusion detection in iomt architecture},
\newblock \bibinfo{journal}{Computer Communications} \bibinfo{volume}{160}
  (\bibinfo{year}{2020}) \bibinfo{pages}{139--149}.
\bibitem[{Lee et~al.(2021)Lee, Cha, Rathore, and Park}]{lee2021m}
\bibinfo{author}{J.~D. Lee}, \bibinfo{author}{H.~S. Cha},
  \bibinfo{author}{S.~Rathore}, \bibinfo{author}{J.~H. Park},
\newblock \bibinfo{title}{M-idm: A multi-classification based intrusion
  detection model in healthcare iot},
\newblock \bibinfo{journal}{Computers, Materials and Continua}
  \bibinfo{volume}{67} (\bibinfo{year}{2021}) \bibinfo{pages}{1537--1553}.
\bibitem[{Salemi et~al.(2021)Salemi, Rostami, Talatian-Azad, and
  Khosravi}]{salemi2021leaesn}
\bibinfo{author}{H.~Salemi}, \bibinfo{author}{H.~Rostami},
  \bibinfo{author}{S.~Talatian-Azad}, \bibinfo{author}{M.~R. Khosravi},
\newblock \bibinfo{title}{Leaesn: Predicting ddos attack in healthcare systems
  based on lyapunov exponent analysis and echo state neural networks},
\newblock \bibinfo{journal}{Multimedia Tools and Applications}
  (\bibinfo{year}{2021}) \bibinfo{pages}{1--22}.
\bibitem[{{MIT Lincoln Laboratory}(1998)}]{MITLincolnLaboratory1998}
\bibinfo{author}{{MIT Lincoln Laboratory}}, \bibinfo{title}{{1998 DARPA
  Intrusion Detection Evaluation Dataset | MIT Lincoln Laboratory}},
  \bibinfo{year}{1998}. \URLprefix
  \url{https://www.ll.mit.edu/r-d/datasets/1998-darpa-intrusion-detection-evaluation-dataset},
  \bibinfo{note}{(Accessed Oct. 07, 2021)}.
\bibitem[{Thamilarasu et~al.(2020)Thamilarasu, Odesile, and
  Hoang}]{thamilarasu2020intrusion}
\bibinfo{author}{G.~Thamilarasu}, \bibinfo{author}{A.~Odesile},
  \bibinfo{author}{A.~Hoang},
\newblock \bibinfo{title}{An intrusion detection system for internet of medical
  things},
\newblock \bibinfo{journal}{IEEE Access} \bibinfo{volume}{8}
  (\bibinfo{year}{2020}) \bibinfo{pages}{181560--181576}.
\bibitem[{Begli et~al.(2019)Begli, Derakhshan, and
  Karimipour}]{begli2019layered}
\bibinfo{author}{M.~Begli}, \bibinfo{author}{F.~Derakhshan},
  \bibinfo{author}{H.~Karimipour},
\newblock \bibinfo{title}{A layered intrusion detection system for critical
  infrastructure using machine learning},
\newblock in: \bibinfo{booktitle}{2019 IEEE 7th International Conference on
  Smart Energy Grid Engineering (SEGE)}, \bibinfo{organization}{IEEE},
  \bibinfo{year}{2019}, pp. \bibinfo{pages}{120--124}.
\bibitem[{{M. Tavallaee} and Ghorbani(2009)}]{KDD22CONF}
\bibinfo{author}{E.~B. {M. Tavallaee}}, \bibinfo{author}{A.~Ghorbani},
  \bibinfo{title}{{Detailed Analysis of the KDD CUP 99 Data Set}},
  \bibinfo{year}{2009}. \URLprefix
  \url{https://www.unb.ca/cic/datasets/nsl.html}, \bibinfo{note}{(Accessed Oct.
  07, 2021)}.
\bibitem[{Alrashdi et~al.(2019)Alrashdi, Alqazzaz, Alharthi, Aloufi, Zohdy, and
  Ming}]{alrashdi2019fbad}
\bibinfo{author}{I.~Alrashdi}, \bibinfo{author}{A.~Alqazzaz},
  \bibinfo{author}{R.~Alharthi}, \bibinfo{author}{E.~Aloufi},
  \bibinfo{author}{M.~A. Zohdy}, \bibinfo{author}{H.~Ming},
\newblock \bibinfo{title}{Fbad: Fog-based attack detection for iot healthcare
  in smart cities},
\newblock in: \bibinfo{booktitle}{2019 IEEE 10th Annual Ubiquitous Computing,
  Electronics \& Mobile Communication Conference (UEMCON)},
  \bibinfo{organization}{IEEE}, \bibinfo{year}{2019}, pp.
  \bibinfo{pages}{0515--0522}.
\bibitem[{{Nour Moustafa}(2019)}]{NourMoustafa2019}
\bibinfo{author}{{Nour Moustafa}}, \bibinfo{title}{{TON{\_}IoT Datasets for
  cybersecurity applications based artificial intelligence}},
  \bibinfo{year}{2019}. \URLprefix
  \url{https://research.unsw.edu.au/projects/toniot-datasets},
  \bibinfo{note}{(Accessed Oct. 07, 2021)}.
\bibitem[{Gupta et~al.(2022)Gupta, Salman, Ghubaish, Unal, Al-Ali, and
  Jain}]{gupta2022cybersecurity}
\bibinfo{author}{L.~Gupta}, \bibinfo{author}{T.~Salman},
  \bibinfo{author}{A.~Ghubaish}, \bibinfo{author}{D.~Unal},
  \bibinfo{author}{A.~K. Al-Ali}, \bibinfo{author}{R.~Jain},
\newblock \bibinfo{title}{Cybersecurity of multi-cloud healthcare systems: A
  hierarchical deep learning approach},
\newblock \bibinfo{journal}{Applied Soft Computing} \bibinfo{volume}{118}
  (\bibinfo{year}{2022}) \bibinfo{pages}{108439}.
\bibitem[{Hameed et~al.(2022)Hameed, Selamat, Latiff, Razak, and
  Krejcar}]{hameed2022whte}
\bibinfo{author}{S.~S. Hameed}, \bibinfo{author}{A.~Selamat},
  \bibinfo{author}{L.~A. Latiff}, \bibinfo{author}{S.~A. Razak},
  \bibinfo{author}{O.~Krejcar},
\newblock \bibinfo{title}{Whte: Weighted hoeffding tree ensemble for network
  attack detection at fog-iomt},
\newblock in: \bibinfo{booktitle}{International Conference on Industrial,
  Engineering and Other Applications of Applied Intelligent Systems},
  \bibinfo{organization}{Springer}, \bibinfo{year}{2022}, pp.
  \bibinfo{pages}{485--496}.
\bibitem[{Wahab et~al.(2022)Wahab, Zhao, Javeed, Al-Adhaileh, Almaaytah, Khan,
  Saeed, and Kumar~Shah}]{wahab2022ai}
\bibinfo{author}{F.~Wahab}, \bibinfo{author}{Y.~Zhao},
  \bibinfo{author}{D.~Javeed}, \bibinfo{author}{M.~H. Al-Adhaileh},
  \bibinfo{author}{S.~A. Almaaytah}, \bibinfo{author}{W.~Khan},
  \bibinfo{author}{M.~S. Saeed}, \bibinfo{author}{R.~Kumar~Shah},
\newblock \bibinfo{title}{An ai-driven hybrid framework for intrusion detection
  in iot-enabled e-health},
\newblock \bibinfo{journal}{Computational Intelligence and Neuroscience}
  \bibinfo{volume}{2022} (\bibinfo{year}{2022}).
\bibitem[{Sharafaldin et~al.(2019)Sharafaldin, Lashkari, Hakak, and
  Ghorbani}]{sharafaldin2019developing}
\bibinfo{author}{I.~Sharafaldin}, \bibinfo{author}{A.~H. Lashkari},
  \bibinfo{author}{S.~Hakak}, \bibinfo{author}{A.~A. Ghorbani},
\newblock \bibinfo{title}{Developing realistic distributed denial of service
  (ddos) attack dataset and taxonomy},
\newblock in: \bibinfo{booktitle}{2019 International Carnahan Conference on
  Security Technology (ICCST)}, \bibinfo{organization}{IEEE},
  \bibinfo{year}{2019}, pp. \bibinfo{pages}{1--8}.
\bibitem[{Schneble and Thamilarasu(2019)}]{schneble2019attack}
\bibinfo{author}{W.~Schneble}, \bibinfo{author}{G.~Thamilarasu},
\newblock \bibinfo{title}{Attack detection using federated learning in medical
  cyber-physical systems},
\newblock in: \bibinfo{booktitle}{Proceedings of the 28th International
  Conference on Computer Communications and Networks (ICCCN), Valencia, Spain},
  volume~\bibinfo{volume}{29}, \bibinfo{year}{2019}.
\bibitem[{Johnson et~al.(2016)Johnson, Pollard, Shen, Lehman, Feng, Ghassemi,
  Moody, Szolovits, {Anthony Celi}, and Mark}]{Johnson2016}
\bibinfo{author}{A.~E. Johnson}, \bibinfo{author}{T.~J. Pollard},
  \bibinfo{author}{L.~Shen}, \bibinfo{author}{L.-w.~H. Lehman},
  \bibinfo{author}{M.~Feng}, \bibinfo{author}{M.~Ghassemi},
  \bibinfo{author}{B.~Moody}, \bibinfo{author}{P.~Szolovits},
  \bibinfo{author}{L.~{Anthony Celi}}, \bibinfo{author}{R.~G. Mark},
\newblock \bibinfo{title}{{MIMIC-III, a freely accessible critical care
  database}},
\newblock \bibinfo{journal}{Scientific Data 2016 3:1} \bibinfo{volume}{3}
  (\bibinfo{year}{2016}) \bibinfo{pages}{1--9}. \bibinfo{note}{(Accessed Oct.
  07, 2021)}.
\bibitem[{Singh et~al.(2022)Singh, Gaba, Kaur, Hedabou, and
  Gurtov}]{singh2022dew}
\bibinfo{author}{P.~Singh}, \bibinfo{author}{G.~S. Gaba},
  \bibinfo{author}{A.~Kaur}, \bibinfo{author}{M.~Hedabou},
  \bibinfo{author}{A.~Gurtov},
\newblock \bibinfo{title}{Dew-cloud-based hierarchical federated learning for
  intrusion detection in iomt},
\newblock \bibinfo{journal}{IEEE Journal of Biomedical and Health Informatics}
  (\bibinfo{year}{2022}).
\bibitem[{Khan et~al.(2022)Khan, Moustafa, Razzak, Tanveer, Pi, Pan, and
  Ali}]{khan2022xsru}
\bibinfo{author}{I.~A. Khan}, \bibinfo{author}{N.~Moustafa},
  \bibinfo{author}{I.~Razzak}, \bibinfo{author}{M.~Tanveer},
  \bibinfo{author}{D.~Pi}, \bibinfo{author}{Y.~Pan}, \bibinfo{author}{B.~S.
  Ali},
\newblock \bibinfo{title}{Xsru-iomt: Explainable simple recurrent units for
  threat detection in internet of medical things networks},
\newblock \bibinfo{journal}{Future Generation Computer Systems}
  \bibinfo{volume}{127} (\bibinfo{year}{2022}) \bibinfo{pages}{181--193}.
\bibitem[{Hady et~al.(2020)Hady, Ghubaish, Salman, Unal, and
  Jain}]{hady2020intrusion}
\bibinfo{author}{A.~A. Hady}, \bibinfo{author}{A.~Ghubaish},
  \bibinfo{author}{T.~Salman}, \bibinfo{author}{D.~Unal},
  \bibinfo{author}{R.~Jain},
\newblock \bibinfo{title}{Intrusion detection system for healthcare systems
  using medical and network data: A comparison study},
\newblock \bibinfo{journal}{IEEE Access} \bibinfo{volume}{8}
  (\bibinfo{year}{2020}) \bibinfo{pages}{106576--106584}.
\bibitem[{{A. A. Hady, A. Ghubaish, T. Salman} and Jain(2019)}]{DTST22CONF}
\bibinfo{author}{D.~U. {A. A. Hady, A. Ghubaish, T. Salman}},
  \bibinfo{author}{R.~Jain}, \bibinfo{title}{{WUSTL EHMS 2020 Dataset for
  Internet of Medical Things (IoMT) Cybersecurity Research}},
  \bibinfo{year}{2019}. \URLprefix
  \url{https://www.cse.wustl.edu/{~}jain/ehms/index.html},
  \bibinfo{note}{(Accessed Oct. 07, 2021)}.
\bibitem[{{Peter Garrett and Joshua
  Seidman}(2011)}]{PeterGarrettandJoshuaSeidman2011}
\bibinfo{author}{{Peter Garrett and Joshua Seidman}}, \bibinfo{title}{{EMR vs
  EHR – What is the Difference? - Health IT Buzz}}, \bibinfo{year}{2011}.
  \URLprefix
  \url{https://www.healthit.gov/buzz-blog/electronic-health-and-medical-records/emr-vs-ehr-difference},
  \bibinfo{note}{(Accessed Oct. 07, 2021)}.
\bibitem[{{Cascio, Teresa}(2021)}]{CascioTeresa}
\bibinfo{author}{T.~{Cascio, Teresa}}, \bibinfo{title}{{electronic health
  record | Description, Implementation, {\&} Issues | Britannica}},
  \bibinfo{year}{2021}. \URLprefix
  \url{https://www.britannica.com/topic/electronic-health-record},
  \bibinfo{note}{(Accessed Oct. 07, 2021)}.
\bibitem[{Boxwala et~al.(2011)Boxwala, Kim, Grillo, and
  Ohno-Machado}]{boxwala2011using}
\bibinfo{author}{A.~A. Boxwala}, \bibinfo{author}{J.~Kim},
  \bibinfo{author}{J.~M. Grillo}, \bibinfo{author}{L.~Ohno-Machado},
\newblock \bibinfo{title}{Using statistical and machine learning to help
  institutions detect suspicious access to electronic health records},
\newblock \bibinfo{journal}{Journal of the American Medical Informatics
  Association} \bibinfo{volume}{18} (\bibinfo{year}{2011})
  \bibinfo{pages}{498--505}.
\bibitem[{Menon et~al.(2014)Menon, Jiang, Kim, Vaidya, and
  Ohno-Machado}]{menon2014detecting}
\bibinfo{author}{A.~K. Menon}, \bibinfo{author}{X.~Jiang},
  \bibinfo{author}{J.~Kim}, \bibinfo{author}{J.~Vaidya},
  \bibinfo{author}{L.~Ohno-Machado},
\newblock \bibinfo{title}{Detecting inappropriate access to electronic health
  records using collaborative filtering},
\newblock \bibinfo{journal}{Machine learning} \bibinfo{volume}{95}
  (\bibinfo{year}{2014}) \bibinfo{pages}{87--101}.
\bibitem[{Ama(2021)}]{Amazon_Access_Data_Competition}
\bibinfo{title}{Amazon access data competition}, \bibinfo{year}{2021}.
  \URLprefix \url{https://sites.google.com/site/amazonaccessdatacompetition/},
  \bibinfo{note}{(Accessed Oct. 07, 2021)}.
\bibitem[{Malin and Bradley(2014)}]{Malin2014}
\bibinfo{author}{Y.~C. Malin}, \bibinfo{author}{Bradley},
\newblock \bibinfo{title}{{Detection of Anomalous Insiders in Collaborative
  Environments via Relational Analysis of Access Logs}},
\newblock \bibinfo{journal}{Bone} \bibinfo{volume}{23} (\bibinfo{year}{2014})
  \bibinfo{pages}{1--7}.
\bibitem[{Marwan et~al.(2018)Marwan, Kartit, and Ouahmane}]{marwan2018security}
\bibinfo{author}{M.~Marwan}, \bibinfo{author}{A.~Kartit},
  \bibinfo{author}{H.~Ouahmane},
\newblock \bibinfo{title}{Security enhancement in healthcare cloud using
  machine learning},
\newblock \bibinfo{journal}{Procedia Computer Science} \bibinfo{volume}{127}
  (\bibinfo{year}{2018}) \bibinfo{pages}{388--397}.
\bibitem[{Sicuranza and Paragliola(2020)}]{sicuranza2020ensuring}
\bibinfo{author}{M.~Sicuranza}, \bibinfo{author}{G.~Paragliola},
  \bibinfo{title}{Ensuring electronic health record cyber-security through an
  hybrid intrusion detection system}, \bibinfo{year}{2020}.
\bibitem[{Synthetichealth(2021)}]{synthetichealth}
\bibinfo{author}{Synthetichealth}, \bibinfo{title}{Synthetichealth/synthea:
  Synthetic patient population simulator}, \bibinfo{year}{2021}. \URLprefix
  \url{https://github.com/synthetichealth/synthea}, \bibinfo{note}{(Accessed
  Oct. 07, 2021)}.
\bibitem[{Newaz et~al.(2020)Newaz, Haque, Sikder, Rahman, and
  Uluagac}]{Newaz2020a}
\bibinfo{author}{A.~I. Newaz}, \bibinfo{author}{N.~I. Haque},
  \bibinfo{author}{A.~K. Sikder}, \bibinfo{author}{M.~A. Rahman},
  \bibinfo{author}{A.~S. Uluagac},
\newblock \bibinfo{title}{{Adversarial Attacks to Machine Learning-Based Smart
  Healthcare Systems}},
\newblock \bibinfo{journal}{2020 IEEE Global Communications Conference,
  GLOBECOM 2020 - Proceedings}  (\bibinfo{year}{2020}).

\end{thebibliography}



\bio{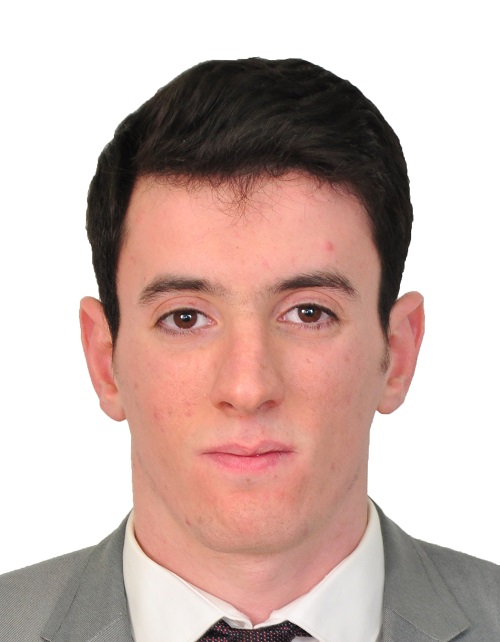}
Ayoub Si-Ahmed obtained his master’s degree in Security of information systems in 2019 at the University of Saad Dahleb Blida, where he finished the major of his promotion. Currently, he is a doctoral student at the same university where he got the post after a competition where he was ranked first. He is part of a research project PRFU (Projets de Recherche Formation-Universitaire) named Security of social networks. He also worked as a consultant in computer security for three years at the PROXYLAN branch of a research center called CERIST. His research interests include ML/DL, IoT systems, and cybersecurity.
\endbio

\bio{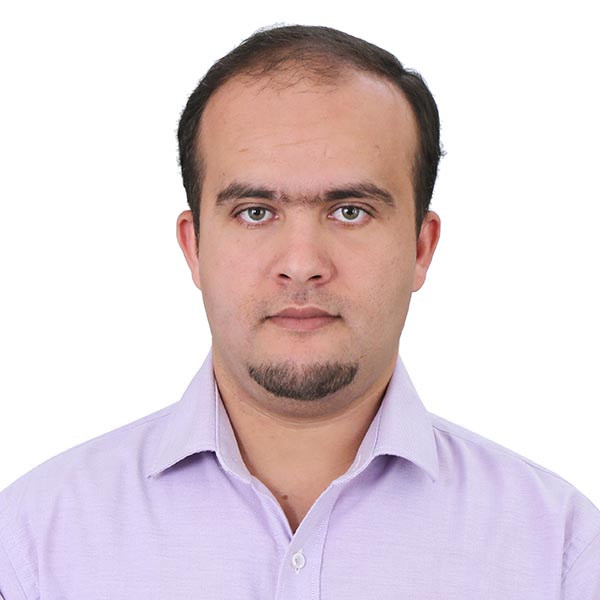}
Mohammed Ali Al-Garadi achieved his PhD from the University of Malaya, Malaysia, in 2017, where he obtained many national and international awards. He has published in many peer-reviewed journals and conferences. He has served as a reviewer in many journals, including IEEE Communications Magazine, IEEE Transactions on Knowledge and Data Engineering, IEEE Access, Future Generation Computer Systems, Computers \& Electrical Engineering, and Journal of Network and Computer Applications. His research focuses on big data analytics, IoT systems, cybersecurity, AI for health care, ML and NLP for medical data.
\endbio

\bio {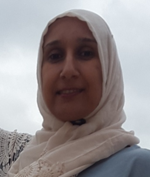}
 Narhimene Boustia received the PhD degree in computer science from USTHB Algiers in 2011. She is Professor and searcher at Blida 1 university. Her research focuses on security of information system, access control and knowledge management.
\endbio

\end{document}